\pdfoutput=1
\documentclass[acmsmall, nonacm, authorversion, noccs]{acmart}
\usepackage{ebutf8}
\usepackage{import}
\usepackage{xspace}
\usepackage{listings}
\usepackage{yade}
\usepackage{tikz-cd}
\usepackage{tikz-cats}
\usepackage{enumitem}
\usepackage[capitalise]{cleveref}

\definecolor{DarkViolet}{rgb}{.58,0,.828}
\definecolor{DarkBlue}{rgb}{0,0,.545}

\newcommand{\plan}[1]{\textcolor{blue}{\textbf{Plan:} #1}\PackageWarning{todo}{todo}}

\newcommand{\todot}[1]{\textcolor{red}{\textbf{Todo:} #1}\PackageWarning{todo}{todo}}

\renewcommand{\plan}[1]{}

\renewcommand{\todot}[1]{}

\usepackage [xcolor,hyperref,notion,electronic]{knowledge}
\knowledge{notion}
  | monoidal category
  | monoidal categories

\knowledge{notion}
  | monoid
  | monoids

\knowledge{notion}
  | module
  | modules

\knowledge{notion}
  | module morphism
  | module morphisms

\knowledge{notion}
 | module signature
 | module signatures

\knowledge{notion}
 | model
 | models

\knowledge{notion}
 | inserter diagram
 | inserter diagrams

\knowledge{notion}
 | vertical inserter
 | vertical inserters

\knowledge{notion}
 | binding-friendly signature
 | binding-friendly signatures

\knowledge{notion}
 | binding-friendly@monoidal
 | binding-friendly monoidal
 | Binding-friendly monoidal

\knowledge{notion}
 | intersectional

\knowledge{notion}
 | finitary@debruijn

\knowledge{notion}
 | finite support

\knowledge{notion}
 | coreflection
 | Coreflection

\knowledge{notion}
 | de Bruijn monad

\knowledge{notion}
 | binding arity

\knowledge{notion}
 | binding signature
 | binding signatures

\knowledge{notion}
 | simply-typed binding signature
 | simply-typed binding signatures

\knowledge{notion}
 | simply-typed binding arity

\knowledge{notion}
 | extended models
 | extended model

\knowledge{notion}
 | derivative
 | derivation

\usepackage{lipsum}

\knowledgenewrobustcmd\VInsDiag{\cmdkl{\mathbf{VInsDiag}}}
\knowledgenewrobustcmd\VIns{\cmdkl{\mathbf{VIns}}}
\knowledgenewrobustcmd\CstDiag{\cmdkl{\mathbf{CstDiag}}}
\knowledgenewrobustcmd\VInsProj{\cmdkl{\mathbf{VInsProj}}}

\knowledgenewrobustcmd\ExtModels[1]{\cmdkl{\mathrm{ExtModels}({#1})}}
\knowledgenewrobustcmd\PModDiag{\cmdkl{\mathbf{PModDiag}}}
\knowledgenewrobustcmd\ModSig{\cmdkl{\mathbf{ModSig}}}
\knowledgenewrobustcmd\ModSigps{\cmdkl{\mathbf{ModSig}_{\mathrm{ps}}}}
\knowledgenewrobustcmd\ModSigDiag{\cmdkl{\mathbf{ModSigDiag}}}

\knowledgenewrobustcmd\Sem[1]{\cmdkl{\mathrm{Sem}_{#1}}}
\knowledgenewrobustcmd\MonCat{\cmdkl{\mathbf{MonCat}}}
\knowledgenewrobustcmd\Model{\cmdkl{\mathbf{Model}}}
\knowledgenewrobustcmd\STSigCat{\cmdkl{\STSig}}
\knowledgenewrobustcmd\BSigModel[1]{\cmdkl{\mathbf{BSigModel}_{#1}}}
\knowledgenewrobustcmd\STSigModel{\cmdkl{\mathbf{STSigModel}}}
\knowledgenewrobustcmd\BindingModelStar[1]{\cmdkl{\mathbf{BindModel}^*_{#1}}}
\knowledgenewrobustcmd\BindMonCat{\cmdkl{\mathbf{BindMonCat}}}
\knowledgenewrobustcmd\BindMonCatStar{\cmdkl{\mathbf{BindMonCat}^*}}
\knowledgenewrobustcmd\BN{\cmdkl{𝐁ℕ}} 
\knowledgenewrobustcmd{\F}{\cmdkl{\mathbb{F}}}
\knowledgenewrobustcmd{\yoneda}{\cmdkl{\mathbf{y}}}

\newcommand{\cat}[1]{\mathsf{#1}\xspace}
\newcommand{\Cat}{\cat{Cat}}

\newcommand{\Mod}{\cat{Mod}}
\newcommand{\Mon}{\cat{Mon}}
\DeclareMathOperator{\Lan}{Lan}
\DeclareMathOperator{\Ran}{Ran}

\newcommand{\C}{\cat{C}}
\newcommand{\D}{\cat{D}}
\newcommand{\LC}{\mathsf{LC}}
\newcommand{\abs}{\mathsf{abs}}
\newcommand{\app}{\mathsf{app}}

\newcommand{\option}{\mathsf{option}}
\newcommand{\op}{\mathsf{op}}
\newcommand{\Kl}{\mathrm{Kl}}

\newcommand{\Set}{\mathbf{Set}}
\newcommand{\doublerightarrow}[3]{\ar[#1,shift left=.75ex,"#2"]\ar[#1,shift right=.75ex,swap,"#3"]}

\newcommand{\NicePSigStrength}{\mathrm{PSigStr}}
\newcommand{\STSig}{\mathrm{STSig}}

\newcommand{\id}{\mathrm{id}}
\DeclareMathOperator{\Id}{Id}

\newcommand{\tcell}{\ensuremath{\Rightarrow}}
\newcommand{\vcomp}{\ensuremath{\bullet}}
\newcommand{\hcomp}{\ensuremath{\circ}}

\usepackage{mathtools}
\usepackage{thm-restate}

\AtEndPreamble{%
\theoremstyle{acmdefinition}
\newtheorem{remark}[theorem]{Remark}
\newtheorem{notation}[theorem]{Notation}
\theoremstyle{acmplain}
\newtheorem*{theorem*}{Theorem}
}

\usepackage{lstcoq}

\begin{document}

\title{2-Functoriality of Initial Semantics, and Applications}

\author{Benedikt Ahrens}
\orcid{0000-0002-6786-4538}
\affiliation{%
  \institution{Delft University of Technology}
  \city{Delft}
  \country{Netherlands}
}
\email{B.P.Ahrens@tudelft.nl}

\author{Ambroise Lafont}
\orcid{0000-0002-9299-641X}
\affiliation{%
  \institution{LIX, Ecole Polytechnique}
  \city{Palaiseau}
  \country{France}
}
\affiliation{%
  \institution{Inria}
  \city{Palaiseau}
  \country{France}
}
\email{ambroise.lafont@lix.polytechnique.fr}

\author{Thomas Lamiaux}
\orcid{0000-0002-7318-5814}
\affiliation{%
  \institution{Nantes University}
  \city{Nantes}
  \country{France}
}
\affiliation{%
  \institution{Inria}
  \city{Nantes}
  \country{France}
}
\email{thomas.lamiaux@etu.univ-nantes.fr}

\begin{abstract}
  Initial semantics aims to model inductive structures and their properties, and
  to provide them with recursion principles respecting these properties.
  An ubiquitous example is the \texttt{fold} operator for lists.
  We are concerned with initial semantics that model languages with variable
  binding and their substitution structure, and that provide
  substitution-safe recursion principles.

  There are different approaches to implementing languages with variable binding
  depending on the choice of representation for contexts and free variables,
  such as unscoped syntax, or well-scoped syntax with finite or infinite contexts.
  Abstractly, each approach corresponds to choosing a different monoidal category
  to model contexts and binding, each choice yielding a different notion of
  ``model'' for the same abstract specification (or ``signature'').

  In this work, we provide tools to compare and relate the models obtained from
  a signature for different choices of monoidal category.
  We do so by showing that initial semantics naturally has a 2-categorical
  structure when parametrized by the monoidal category modeling contexts.
  We thus can relate models obtained from different choices of
  monoidal categories provided the monoidal categories themselves are related.
  In particular, we use our results to relate the models of the different
  implementation --- de Bruijn vs locally nameless, finite vs infinite contexts ---,
  and to provide a generalized recursion principle for simply-typed syntax.
\end{abstract}



\maketitle


\section{Introduction}
\label{sec:introduction}

When studying abstract syntax and initial semantics for it, a mathematical
modelling of contexts and variable binding needs to be chosen.
This choice, while seemingly insignificant, determines the remainder of the
theory, and shapes the recursion principle obtained from the initiality property.
In this work, we provide an abstract framework formalizing the dependency of the
relating different ways of representing contexts and variable binding.

In the remainder of the introduction, we provide a gentle introduction to
initial semantics (in \cref{subsec:is-binding}) and to the particular challenge
of modelling substitution (in \cref{subsec:is-subst}).

\subsection{Initial Semantics for Variable Binding}
\label{subsec:is-binding}

Initial semantics \cite{Lawvere:thesis,goguen1974initial,InitialSemantics} aims to characterize an inductive structure --- e.g., the
abstract syntax of a language --- as an initial object in a suitable category.
The property of being an initial object amounts to a recursion principle that
can be used to specify maps --- e.g., translations of the abstract syntax ---
into objects of the same category.
An ubiquitous instance of this principle is Haskell's \texttt{fold} operator,
which exploits the fact that the type \texttt{[a]} of lists over base type
\texttt{a}, together with the constructors \texttt{[] :: [a]} and \texttt{(:) ::
a -> [a] -> [a]}, is the initial object in a suitable category of
``list-algebras''.
Specifically, a list-algebra consists of a type \texttt{b} (the ``carrier'') and
two functions, \texttt{nil :: b} and \texttt{cons :: a -> b -> b}, and a
morphism between two list-algebras is a function on the underlying carriers that
preserves \texttt{nil} and \texttt{cons} in a suitable way.

In this work, we are concerned with more complicated languages featuring
\emph{variable binding}, such as the lambda calculus.
When implementing and formalizing languages such languages, the question of how to model
variable binding and substitution arises.
The chosen answer to this question leads to different notions of renaming and
substitution, and to different categories of ``algebras'' where, in particular,
the types of the ``carriers'' differ.

Consider the implementation\footnote{
These examples are given in Rocq syntax, but that is irrelevant in the remainder
of the paper; in particular, no knowledge of Rocq is required.
} of the untyped lambda calculus using natural De Bruijn variables to represent
variables as given in \cref{code:ULC-unscoped}.
This definition is also known as \emph{unscoped} syntax, since free and bound variables
are not directly specified by the definition of the syntax, but require a separate definition on top of it.

\begin{center}
\begin{lstlisting}[language={coq}, label={code:ULC-unscoped}, caption={Unscoped Lambda Calculus},
                    captionpos=b, xleftmargin=.35\textwidth, xrightmargin=.35\textwidth]
Inductive LC : Set :=
| Var : Nat -> LC
| App : LC -> LC -> LC
| Abs : LC -> LC
\end{lstlisting}
\end{center}

Taking an approach analogous to the list-algebras as before, we could define a
LC-algebra to be a carrier set $X : \Set$ together with suitable functions
$var : \mathbb{N} \to X$, $app : X \to X \to X$, and $abs : X \to X$ compatible
with substitution.
With a suitable notion of morphism of LC-algebras, the abstract syntax of the
lambda calculus constitutes the initial LC-algebra.

Another possible formalization of the lambda calculus is to use \emph{well-scoped}
syntax, also known as \emph{intrinsic} syntax, with finite or infinite contexts as
given in \cref{code:ULC-wscoped-finite,code:ULC-wscoped-infinite}.
In this implementation, the context of variables is explicit and directly part
of the typing of the meta-language.
The advantage of this approach is that all terms are well-scoped by construction.

\medskip
\noindent
\begin{minipage}{.45\textwidth}
\begin{lstlisting}[language=coq, label={code:ULC-wscoped-finite},
                    caption={Well-scoped Lambda Calculus with Finite Contexts}]
Inductive LC : Nat -> Set :=
| Var : forall n : Nat, Fin n -> LC n
| App : forall n : Nat, LC n -> LC n -> LC n
| Abs : forall n : Nat, LC (n + 1) -> LC n
\end{lstlisting}
\end{minipage}\hfill
\begin{minipage}{.45\textwidth}
\begin{lstlisting}[language=coq, label={code:ULC-wscoped-infinite},
                    caption={Well-scoped Lambda Calculus with Infinite Contexts}]
Inductive LC : Set -> Set :=
| Var : forall X : Set, X -> LC X
| App : forall X : Set, LC X -> LC X -> LC X
| Abs : forall X : Set, LC (X + 1) -> LC X
\end{lstlisting}
\end{minipage}
\medskip

A LC-algebra for well-scoped syntax with infinite contexts consists of a functor
$F : \Set \to \Set$ associating well-scoped terms to contexts, together with
natural transformations $var : 1 \to F$, $app : F \times F \to F$, and $abs : F
\circ \mathsf{option} \to F$ modeling the constructors.
Here $\mathsf{option} : \Set \to \Set$ maps a set $X$ to the set $X+1$.
With a suitable notion of morphism of LC-algebras, the abstract syntax of the
lambda calculus constitutes the initial LC-algebra.
A LC-algebra for well-scoped syntax with finite contexts is similar except
that it uses a functor $F : \F \to \Set$, where \AP $\intro*\F$ is the category
of finite cardinals (whose objects are natural numbers).

All the mentioned approaches have been used to prove initial semantics results
for syntax with variable binding. Specifically, \citet{DBLP:conf/lics/FiorePT99} used the category $[\F, \Set]$ of functors from
the category of finite cardinals and any map between them to sets;  \citet{DBLP:journals/iandc/HirschowitzM10} used the category $[\Set,
\Set]$, thus modelling languages as monads on sets; \citet{dblmcs} studied abstract syntax using De
Bruijn variables.

All these semantic frameworks account for substitution from the start using suitable monoidal structures, as we will see in the next subsection. This is in contrast with the approach to variable binding based on nominal sets~\cite{PittsAM:newaas}, which we do not investigate in this paper.

\subsection{Initial Semantics for Substitution}
\label{subsec:is-subst}

Modeling the untyped lambda calculus as an initial algebra only captures the abstract syntax.
It does not model substitution and how (variable binding) constructors interact
with it, even though that is the core characteristic of languages with variable binding.

\citet{Linton:equational-cats} discovered that simultaneous substitution
 equips the lambda
 calculus as given in \cref{code:ULC-wscoped-infinite} with the structure of a monad.
 This insight was rediscovered later in different contexts, see, for instance,
 the work by
 \citet{DBLP:journals/scp/BellegardeH94},
 \citet{bird_paterson_1999}, and
\citet{DBLP:conf/csl/AltenkirchR99}.
Monads are equivalently monoids for composition: in this respect, the inclusion
of variables into terms provides the unit of the monad, and the flattening of
terms with terms as variables constitutes the monad multiplication.
This is also equivalent to monads in terms of ``extension systems''  whose \texttt{bind}
operation directly corresponds to simultaneous substitution;
see, for instance, Section 3, Exercise 12 of the book by \citet{manes:algebraic-theories}.

We can extend the definition of LC-algebra, in this
approach, as follows: the carrier is a monad (not just a functor) on the
category of sets, and the operations $app$ and $abs$ should preserve the
substitution structure of the monad in a suitable way.
Expressed concisely, the carrier of a LC-algebra is a monoid in the monoidal
category $([\Set,\Set], \circ, I)$, where the monoidal product is given by
composition $G \circ F$ of endofunctors, with the unit $I$ being the identity
functor.

The other approaches can be upgraded, in a similar way, to incorporate
substitution for different choices of the monoidal category.
In the finite-context approach, a carrier can be defined to be a monoid in the
monoidal category $([\F,\Set], \otimes, J)$, for a suitable monoidal product
$\otimes$ (involving a left Kan extension).
In the De Bruijn approach, a carrier can be defined to be a relative monad on
the inclusion $\BN → \Set$ where $\BN$ is the full subcategory of sets
consisting of $ℕ$ as its single object~\cite{dblmcs}.

Abstractly, to all of these different ways of modelling syntax corresponds a
suitable \emph{monoidal category} modelling contexts, variable binding, and
substitution.
Given a notion of signature that specifies the constructors of a language, this
choice of a monoidal category leads to a corresponding notion of ``algebra'' or
``model'' of a signature.
These different notions of model differ in the carriers, modelling the
language-independent details on how contexts are modelled, but contain the same
language-specific information, in particular, the types of the language
constructors.

\subsection{Contributions}
\label{subsec:contributions}

Most lines of work on initial semantics of which we are aware start by fixing a choice
of base monoidal category, e.g. to study a particular class of languages or
semantics like reduction rules, and keep it fixed throughout the paper.
In this paper, we do the opposite and study how to \emph{change} the monoidal
category modelling contexts, and its applications.
Specifically, given two base monoidal categories satisfying a relationship like
an adjunction or an equivalence, how do the resulting categories of models relate?

To do so, we identify that \kl{module signatures}, once parametrized by the
choice of monoidal category, form a 2-category, and that the category of models
can then be computed by a 2-functor.
More specifically:

\begin{enumerate}
  \item We identify that, once parametrized by monoidal categories, \kl{module signatures}
        form a 2-category $\ModSig$.
  \item There is a 2-functor $\ModSig → \Cat$ that computes the category of models
        of a signature (\cref{thm:2-functoriality-models}).
\end{enumerate}

As a consequence, functors, adjunctions, or equivalences, between different kinds
of contexts, lift to functors, adjunctions, or equivalences between the
resulting categories of models.
We then use the abstract machinery to establish old and new concrete results about
the relationship between models of syntax using different monoidal categories.

We first recover, in a general and systematic way, links between the models
of the different implementations of abstract syntax like the lambda calculus mentioned above.

\begin{enumerate}
\item We define a 2-category $\BindMonCat$ of monoidal categories with enough
      structure to interpret any binding signature.
      We then show that given binding signature $S$, there is a 2-functor
      $\Sem{S} : \BindMonCat → \ModSig$ (\cref{prop:param-module-funct}).

\item Using this result, we construct a coreflection between
      the models of well-scoped syntax with finite contexts (using $[\F, \Set]$)
      and the one using infinite contexts (using $[\Set, \Set]$), cf \cref{thm:lifting-models}.
      Hence, we recover a result which previously only had a long proof, fully
      constructed by hand by \citet{ZsidoPhd10}.

\item Similarly, we prove that models of unscoped syntax (using $[\BN,\Set]$)
      and of well-scoped syntax with finite contexts (using $[\F, \Set]$) are
      equivalent once restricted to well-behaved ones (\cref{thm:lifting-models-debruijn}),
      which only had been proven by hand by \citet{HHLM}.
\end{enumerate}

We then study how changing the base monoidal categories enables us to build more
general recursion principles for simply-typed languages, as the one generated by
initiality is by default limited to a fixed type system.

\begin{enumerate}
  \item We define a category $\STSigCat$ of \kl{simply-typed binding signatures}, and show there is a
        1-functor $\STSigModel ∶\STSigCat^{\op} → \ModSig → \Cat$ computing the category of models
        (\cref{def:Models-STSig}).
        As a consequence, a morphism between simply-typed signatures over different
        type systems lifts to a functor between their categories of models.
  \item Building upon $\STSigModel$,
    we recover Ahrens' category of models~\cite{ExtendedInitiality12}
        in \cref{prop:revisiting-ahrens}
        using the \emph{Grothendieck construction}~\cite[Definition 1.10.1]{Jacobs},
        hence providing a new insight on this framework specifically tailored for translating
        across type systems.
        This category  gathers models over different object types in one ``large'' category;
        translations between languages over different types can thus be viewed as morphisms in this category.
\end{enumerate}

\subsection{Synopsis}

In \cref{sec:overview-initial-semantics}, we start by giving a brief introduction
to initial semantics, on which we rely in the remainder of the paper.
We also give a brief introduction to the very few notions of 2-category theory
needed to understand this paper in \cref{sec:overview-2cats}.
Note that we only use \emph{strict} 2-category theory, where axioms about
1-cells are expressed modulo equality, not modulo invertible 2-cells.
We show in \cref{sec:2-cat-perspective} that initial semantics has a
2-categorical structure, and that models can be computed by a 2-functor.
We then define \emph{binding-friendly} monoidal categories in
\cref{sec:binding-friendly-mon-cat} to leverage the 2-functoriality of models to
relate the models of the different implementations of the untyped lambda
calculus in \cref{sec:application-FSet,sec:application-equivalence-debruijn}.
We also leverage 2-functoriality of models to prove a generalized
recursion principle for simply-typed syntax in \cref{sec:app-generalized-rec}.
We discuss related work in \cref{sec:related-work}, and conclude in \cref{sec:conclusion}.

\section{A Short Introduction to Initial Semantics}
\label{sec:overview-initial-semantics}

In this section, we give a short introduction to initial semantics.
It is necessarily terse and incomplete, and we refer to the work by
\citet{lamiaux2025unifiedframeworkinitialsemantics} for details and additional examples.

In this section, we discuss initial semantics in a fixed monoidal category.
Later, in \cref{sec:2-cat-perspective}, we  give these definitions a 2-categorical structure depending on a 2-category of monoidal categories.

\subsection{Abstracting Syntax and Substitution}

Initial semantics requires a base category to model context and variable binding.
To model syntax, it requires more than the structure of a mere category,
it requires a \emph{monoidal} category.
In this section, we briefly recall the basic categorical definitions and sketch how they model syntax and substitution.
For details on monoids, we refer to Chapter VII of the book by \citet{MacLane:cwm}.

  Recall that a \emph{\AP\intro{monoidal category}} \cite[VII.1]{MacLane:cwm} is a tuple $(\C,⊗,I,α,λ,ρ)$, where $\C$
  is a category, $\_ ⊗ \_ : \C × \C \to \C$ is a bifunctor called the monoidal
  product, and $I : \C$ an object called the unit.
  Furthermore, we have natural isormorphisms $α,λ, ρ$ as below -- called the associator, and the
  left and right unitor -- that satisfy the unit axiom and the pentagon axiom.
  \begin{align*}
    α_{X,Y,Z} : (X ⊗ Y) ⊗ Z ≅ X ⊗ (Y ⊗ Z)
    &&
    λ_{X} : I ⊗ X ≅ X
    &&
    ρ_{X} : X ⊗ I ≅ X
  \end{align*}

In the following, we will simply write $(\C,⊗,I)$ or even $\C$ for a monoidal
category, leaving the other components implicit.
The main interests of monoidal categories is that syntax and simultaneous
substitution can be modeled as \emph{monoids} in particular monoidal categories.

  Given a monoidal category $(\C,⊗,I)$, a \emph{\AP\intro{monoid}} on $\C$ \cite[VII.3]{MacLane:cwm} is a tuple
  $(R,μ,η)$ where $R$ is an object of $\C$ and the \emph{multiplication}
  $μ : R ⊗ R → R$ and the \emph{unit} $η : I → R$ are morphisms of $\C$ satisfying the monoid laws.
  This forms a category $\Mon(\C)$ together with morphism of monoids.

  \begin{example}[\cite{DBLP:journals/scp/BellegardeH94,bird_paterson_1999,DBLP:conf/csl/AltenkirchR99}]
    \label{ex:lc-monad}
  The lambda calculus $\LC : \Set \to \Set$ as defined in \cref{code:ULC-wscoped-infinite},
  together with a suitable substitution operation, is a monad on $\Set$, that
  is, a monoid on $([\Set, \Set], \circ, I)$.
\end{example}

Monoids are not enough on their own to fully model languages with variable binding,
as monoids do not capture constructors and their substitution structures.
To do so, on top of monoids, we model constructors as morphisms of modules over monoids.

\begin{definition}[Category of Modules over a Monoid]
  \label{def:modules}
  Given a monoid $R : \Mon(\C)$, a (left) $R$-\AP\intro{module} is a tuple $(M,
  p^M)$ where $M$ is an object of $\C$ and $p^M : M ⊗ R → M$ is
  a morphism of $\C$ called \emph{module substitution} that is compatible
  with the multiplication and the unit of the monoid:
  \begin{align*}
    \begin{tikzcd}[ampersand replacement=\&,column sep=large]
      (M ⊗ R) ⊗ R \ar[r, "\alpha_{M,R,R}"] \ar[d, swap, "p^M ⊗ R"]
        \& M ⊗ (R ⊗ R) \ar[r, "M ⊗ μ"]
        \& M ⊗ R \ar[d, "p^M"] \\
      M ⊗ R \ar[rr, swap, "p^M"]
        \&
        \& M
    \end{tikzcd}
    &&
    \begin{tikzcd}[ampersand replacement=\&]
      M ⊗ I \ar[r, "M ⊗ η"] \ar[dr, swap, "\rho_M"]
        \& M ⊗ R \ar[d, "p^M"] \\
        \& M
        \&
    \end{tikzcd}
  \end{align*}

  Given two modules $(M, p)$ and $(M', p')$ over $R$, a \AP\intro{module morphism} from $(M, p)$ and $(M', p')$ is a morphism $r : M → M'$ of $\C$ commuting with
  the respective module substitutions:
  \begin{equation}\label{eq:module-morphism}
    \begin{tikzcd}
      M ⊗ R \ar[r, "r ⊗ R"] \ar[d, swap, "p"]
        & M' ⊗ R \ar[d, "p'"] \\
      M \ar[r, swap, "r"] & M'
    \end{tikzcd}
  \end{equation}

With composition and identity induced by that of the monoidal category $\C$, modules over $R$ and their morphisms form a category, which we call $\Mod(R)$.
\end{definition}

\begin{definition}
  Any monoid $R$ induces a module over itself, also called $R$.
\end{definition}

\begin{example}
  The functor $X \mapsto \LC(X) \times \LC(X)$ yields a module over the monad $\LC$ of \cref{ex:lc-monad}.
  The functor $X \mapsto \LC(X + 1)$ yields a module over the monad $\LC$ of \cref{ex:lc-monad}.
  For both, the module substitution is given by parallel substitution.

  The constructors $\app : \LC \times \LC \to \LC$ and $\abs : \LC \circ \option
  \to \LC$ are morphisms of $\LC$-modules.
  For both, Diagram \ref{eq:module-morphism} spells out the commutation of $\app$ and $\abs$ with substitution, that is,
  $(\app (M,N))[f] = \app(M[f], N[f])$ and $(\abs(M))[f] = \abs(M[f\uparrow])$, where $f\uparrow$ is a suitable lift of the substitution function $f$, necessary when descending under a variable binder.
\end{example}

\subsection{Signatures and Models}

Initial semantics aims to provide a generic framework for studying syntax,
therefore, it requires a generic notion of signatures and of models.
As constructors are represented by morphisms of modules over monoids,
we would like to take this as our specification.
However, modules are defined over specific monoids, and we need signatures
to specify languages for any choice of monoids.
To express this, we use the total category of modules over monoids.

\begin{definition}[Total category of modules]
  There is a category of modules over monoids, denoted $\Mod(\C)$.
  Its objects are tuples $(R,M)$ where $R : \Mon(C)$ is a monoid, and
  $M : \Mod(R)$ a module over it.
  Its morphisms $(R,M) → (R',M')$ are tuples $(f,r)$ where $f : R → R'$
  is a morphism of monoids and $r : M → f^*M'$ a morphism of $R$-monoids.
  Here, $f^*(M') : \Mod(R)$ denotes the $R$-module with underlying object $M'$
  and module multiplication $M' ⊗ R \xrightarrow{M' ⊗ f} M' ⊗ R' \xrightarrow{p^{M'}} M'$.
\end{definition}

\begin{remark}
  The category $\Mod(C)$ is usually denoted by $\int_{X : \Mon(C)} \Mod(X)$ as
  it is the total category for the 1-functor $\Mod : \Mon(C)^\op → \Cat$.
  In this work, we denote it instead by $\Mod(C)$ for brevity.
\end{remark}

It is then possible to specify languages by functors $\Mon(C) ⟶ \Mod(C)$ that
return a module over its input.

\begin{definition}[Signature]
  A \AP\intro{module signature} is a functor $\Sigma : \Mon(C) ⟶ \Mod(C)$ such that
  $π_1 ∘ \Sigma = \id$, that is $Σ$ is a section of $π_1$.
  In other words, such that the following diagram commute:
  \[
    \begin{tikzcd}
      \Mon(\C) \ar[rr, "\Sigma"] \ar[dr, swap, equal]
        &
        & \Mod(\C) \ar[dl, "π_1"] \\
      & \Mon(\C) &
    \end{tikzcd}
  \]
\end{definition}

\begin{example}[Signature of LC]\label{ex:sig-LC}
  The signature of the lambda calculus, considered in the monoidal category
  $[\Set,\Set]$, maps a monad $R$ to the $R$-module $R \times R + R \circ
  \option$, thus specifying the source of the two domain-specific constructors
  $\app$ and $\abs$.
\end{example}

\begin{definition}\label{def:theta}
  There is a trivial signature $\Theta : \Mon(C) → \Mod(C)$ that maps a monoid
  $(R,η,μ)$ to the module $(R,μ)$.
\end{definition}

\begin{example}[Signature of LC, continued]\label{ex:sig-LC-cont}
  Using the trivial signature \cref{def:theta} to represent a unary operation, and writing $\Theta^{(1)} : R \mapsto R \circ \option$,
  we can write the signature of the lambda calculus from \cref{ex:sig-LC} as $\Theta \times \Theta + \Theta^{(1)}$.
\end{example}

To specify constructors, we need two signatures: one for the input and one for the output.
Since we always return regular terms without additional free variables, and to simplify the
framework, we fix the output signature to be the trivial signature Θ.
Thus, we only need one signature Σ to specify the constructors.
We are now ready to define the models of a module signature.

\begin{definition}[Category of models of a signature]
  A \AP\intro{model} of a \kl{module signature} $\Sigma$ consists of a pair $(R,m)$ of a monoid $R$
  together with a morphism $m : \Sigma(R) \to R$ of $R$-modules.

  A morphism $f : (R,m) \to (R',m')$ of models is a morphism of monoids $f : R \to R'$ that commutes with the module morphisms:
  \[
  \begin{tikzcd}[ampersand replacement=\&]
      \Sigma(R) \ar[r, "r"] \ar[d, swap, "\Sigma(f)"]
        \& R \ar[d, "f"] \\
      f^* \Sigma(R') \ar[r, swap, "f^*r'"]
        \& f^*R'
      \end{tikzcd}
    \]
  Models of $\Sigma$, and their morphisms, form the category $\Model(Σ)$ of models of $Σ$.
\end{definition}

This provides a generic and abstract framework to model syntax and its properties.
Initial semantics is then concerned with proving initiality theorems, that is,
theorems asserting the existence of an initial model, under hopefully simple conditions
on the \kl{monoidal category} and the \kl{module signature} specifying a language.
The existence of this model then ensures the existence of an implementation, and
the initiality provides it with a recursion principle respecting the substitution
structure of the language.
In this work, we are not concerned with initiality theorems, but such a theorem
and a discussion about them can be found, e.g., in a paper by \citet{lamiaux2025unifiedframeworkinitialsemantics}.

\section{Overview of 2-Category Theory}
\label{sec:overview-2cats}

In this section we recall the few notions of 2-categories needed to understand the paper, and state some results used later.
\citet{Companion} provides an extensive introduction to 2-categories, as does \citet[XII.3-4]{MacLane:cwm}.

Note that our 2-categories and related structures are all ``\emph{strict}''.
This means that they satisfy laws about 1-cells up to equality, not only up to (invertible) 2-cell;
see also \cref{rem:2-vs-bi}.

A 2-category is a category \emph{enriched} in the
cartesian category $\Cat$ of categories. Equivalently, a 2-category is a 1-category with a ``system of 2-cells or `maps' which can be composed in two different but commuting categorical ways'', to quote \citet[XII.3]{MacLane:cwm}.
\begin{definition}\label{def:2-cat-2cell}
A 2-category $\C$ consists of
\begin{enumerate}
\item an underlying 1-category; %
\item for each pair of objects $(a,b)$, a \emph{hom-category} $\underline{\C(a,b)}$ whose underlying set of objects is $\C(a,b)$; we denote its composition by $\vcomp$ and draw a morphism $α$ from $f$ to $g$ as follows;
  \[
    \begin{tikzcd}[column sep=large, row sep=large]
    a \ar[r, "f", ""'{name=UU}, bend left=30] \ar[r, "g"', ""{name=VV}, bend right=30]
    &
    b
    \ar[from=UU, to=VV, Rightarrow, "\alpha",outer sep=2pt]
  \end{tikzcd}
\]
\item an associative \emph{horizontal composition} operation mapping morphisms $\alpha$ and $\beta$ in hom-categories $\C(a,b)$ and $\C(b,c)$ (on the left)  to their horizontal composite (on the right):
  \[
    \begin{tikzcd}[column sep=large, row sep=large]
      a \ar[r, "f", ""'{name=UU}, bend left=30] \ar[r, "g"', ""{name=VV}, bend right=30]
      &
      b \ar[r, "f'", ""'{name=WW}, bend left=30] \ar[r, "g'"', ""{name=XX}, bend right=30]
      &
      c
      \ar[from=UU, to=VV, Rightarrow, "\alpha",outer sep=2pt]
      \ar[from=WW, to=XX, Rightarrow, "\beta",outer sep=2pt]
    \end{tikzcd}
    \quad\mapsto\quad
  \begin{tikzcd}[column sep=large, row sep=large]
    a \ar[rr, "f' \circ f", ""'{name=UU}, bend left=30] \ar[rr, "g' \circ g"', ""{name=VV}, bend right=30]
    &
    &
    c
    \ar[from=UU, to=VV, Rightarrow, "\beta \hcomp \alpha",outer sep=2pt]
  \end{tikzcd}
\]

\end{enumerate}
such that the horizontal composition of two identities is the identity, and such that the following equations hold:
$1_{1_b} \hcomp f = f = f \hcomp 1_{1_a}$, and $(\beta' \hcomp \beta) \vcomp (\alpha' \hcomp \alpha) = (\beta' \vcomp \alpha') \hcomp (\beta \vcomp \alpha)$.
Here, $1_{1_x}$ denotes the identity morphism in $\C(x,x)$ on the identity $1_x$ on $x$.
\end{definition}

We refer to the morphisms of the underlying 1-category $\C$ also as $1$-cells or 1-morphisms,
and to the composition of that 1-category as 1-composition.
Similarly, for each pair of objects $a,b$, we refer to the morphisms of the category $\underline{\C(a,b)}$ as $2$-cells or 2-morphisms,
and to the composition of that category as vertical composition.

\begin{example}\label{ex:2-cat-of-cats}
Our prototypical example of 2-category is given by the 2-category of categories, functors, and natural transformations.
This 2-category has, as its collections of objects, all (small) categories.
For given categories $\C$ and $\D$, their hom-category is given by the functor category $[\C,\D]$ of functors from $\C$ to $\D$, and natural transformations between them.
\end{example}

Given a 2-category $\C$, we write $\C_0$ also for its underlying collection of objects;
we write $\C(a,b)$ for the collection of 1-cells or 1-morphisms from $a$ to $b$,
and $\C(f,g)$ for the collection of 2-cells from $f$ to $g$.
We write 1-composition of $f : \C(a,b)$ and $g : \C(b,c)$ as $g \circ f$ or $gf$.
We also write $\alpha f$ for the horizontal composition $\alpha \hcomp 1_f$ --- this is known as ``whiskering'' --- and similar for whiskering in the other order.
A 2-cell $\alpha : f \tcell g$ is an isomorphism if there is a 2-cell $\alpha' : g \tcell f$ such that $\alpha \vcomp \alpha' = 1_g$ and $\alpha' \vcomp \alpha' = 1_f$.

\begin{example}
  Rephrasing \cref{ex:2-cat-of-cats} in terms of \cref{def:2-cat-2cell}, the 2-category of categories has, as underlying 1-category, the 1-category of (small) categories. The 2-cells from functor $F : \C \to \D$ to functor $G : \C \to \D$ are given by the natural transformations $\alpha : F \tcell G$.
\end{example}

\begin{remark}\label{rem:2-vs-bi}
  The 1-composition of a 2-category satisfies the usual categorical laws \emph{strictly}, that is, up to equality.
  This is in contrast to the weaker notion of a \emph{bi}category \cite[XII.6]{MacLane:cwm}, where the 1-categorical laws only hold up to an invertible 2-cell.
  The notion of bicategory is thus more general than that of a 2-category; we do not have any use for that generality and work exclusively with 2-categories.
\end{remark}

We continue the axiomatization of our main example of 2-categories, the 2-category of categories (\cref{ex:2-cat-of-cats}).
In particular, the different notions of maps between categories, in particular, adjunction, coreflection, and equivalence,
directly generalize to 2-categories:

\begin{definition}[Adjunction in a 2-category]
  Let $\C$ be a 2-category.
  An adjunction in $\C$ consists of two 1-morphisms $l ∶ a → b $ and $r ∶ b → c$
  with two 2-cells $ η ∶ \id_c → r ∘ l$ and $ε ∶ l ∘ r → \id_a$, called the unit
  and the counit, subject to the usual triangle equalities $εf \vcomp fη = 1_f$
  and $gε \vcomp ηg = 1_g$.
  We write $l : a ⊣ b : r$ or $l ⊣ r$ to denote this situation.
\end{definition}

\begin{definition}[Coreflection in a 2-category]
  A \AP\intro{coreflection} is an adjunction $l ⊣ r$ such that the unit $η$ is an isomorphism.
\end{definition}

\begin{definition}[Equivalence in a 2-category]
  An equivalence is an adjunction $l ⊣ r$ such that both $η$ and $ε$ are isomorphisms.
\end{definition}

\begin{definition}
  Given a 2-category $\C$ and $f : \C(a,b)$, we define the functor
  \[ f_* : \underline{\C(x,a)} \to \underline{\C(x,b)} \]
  given on objects as $g \mapsto f g$ and on morphisms as $\alpha \mapsto f \alpha$.

  We call $f$ \emph{fully faithful} if $f_*$ is a fully faithful functor.
\end{definition}

\begin{lemma}
  An adjunction $l ⊣ r $ in $\Cat$ is a coreflection if and only if $l$ is full and faithful.
\end{lemma}

We now review \emph{morphisms of 2-categories}, to which we refer to as \emph{2-functors}.
Intuitively, a 2-functor from $\C$ to $\D$ is a 1-functor between the underlying 1-categories that also acts on the 2-cells, preserving their source and target.

\begin{definition}[2-functor]
  Given 2-categories $\C$ and $\D$, a \emph{2-functor} $F : \C \to \D$ from $\C$
  to $\D$ consists of functions on the objects, morphisms, and 2-cells from $\C$
  to $\D$ that preserve source and target; in detail,
  \begin{enumerate}
  \item $F_0 : \C_0 \to \D_0$;
  \item $F_1 : \C(a,b) \to \D(F_0a, F_0b)$;
  \item $F_2 : \C(f,g) \to \D(F_1f, F_1g)$.
  \end{enumerate}
  In practice, we omit the indices, since they will be clear from the context.
  Furthermore, identities and compositions are preserved by the functions on morphisms and 2-cells:
  $F(1_a) = 1_{Fa}$, $F(1_f) = 1_{Ff}$, $F(gf) = (Fg)(Ff)$, and $F(\beta\alpha) = (F\beta)(F\alpha)$.
\end{definition}

The important result for the rest of the paper is the following:

\begin{proposition}\label{prop:2-functors-preserve-relations}
  2-functors preserve adjunctions, coreflections, and equivalences.
\end{proposition}

\section{A 2-Categorical Perspective on Initial Semantics}
\label{sec:2-cat-perspective}

In this section, we show how the category of \kl{models} of a signature can be computed by a
2-functor from the 2-category $\ModSig$ of \kl{module signatures}
 (\cref{thm:2-functoriality-models}).
Leveraging the 2-functoriality of this construction, we can then study changes in the base
monoidal category (\cref{sec:binding-friendly-mon-cat,sec:application-FSet,sec:application-equivalence-debruijn}) and
develop a generalized recursion principle for simply typed syntax
(\cref{sec:app-generalized-rec}).

Signatures can be formulated as particular ``(vertical) inserter diagrams'' (\cref{def:v-ins-diag})
In \cref{subsec:models-as-inserters}, we construct a 2-functor from the 2-category of inserter diagrams to the 2-category $\Cat$;
this 2-functor computes the category of models of an inserter diagram.

In \cref{subsec:models-modules} we construct another 2-functor, from module signatures to vertical inserters.
By composing these two 2-functors, we can then compute the category of models of a module signature, in a 2-functorial way.

\subsection{Categories of Models as Vertical Inserters}
\label{subsec:models-as-inserters}

Here, we first show in \cref{subsubsec:abstract-sigs-mods} how the \kl{module signatures} and \kl{models} reviewed in
\cref{sec:overview-initial-semantics} can be formulated as \kl{inserter diagrams}
and \kl{vertical inserters}, respectively.
We study the 2-categorical structure of inserter diagrams in \cref{subsubsec:2-cat-struct-ins}.
We then use these intermediate notions and their 2-categorical structure, in \cref{subsubsec:models-from-inserter-diags}, to construct a 2-functor that computes the
category of models from an inserter diagram.

\subsubsection{Abstracting Signatures and Models}\label{subsubsec:abstract-sigs-mods}

We show how module signatures are subsumed by ``vertical inserter diagrams'', and their categories of models by inserters of inserter diagrams.

A \kl{module signature} $Σ$ induces the following
diagram with $π_1 ∘ Σ = \id = π_1 ∘ Θ$ :
\[
\begin{tikzcd}
  \Mon(C) \doublerightarrow{r}{Σ}{Θ} & \Mod(C) \ar[r, "π_1"] & \Mon(C)
\end{tikzcd}
\]

\noindent This structure naturally generalizes as inserter diagrams $(Σ,Θ,π_1)$:

\begin{definition}\label{def:v-ins-diag}
  A vertical \AP\intro{inserter diagram} consists of functors $F,G: A → B$
  and a functor $p: B → C$ such that $p ∘ F = p ∘ G$ as below:
  \[
  \begin{tikzcd}
    A \doublerightarrow{r}{F}{G} & B \ar[r, "p"] & C
  \end{tikzcd}
  \]
\end{definition}
In what follows, we may omit the adjective ``vertical'' and simply speak of ``inserter diagrams''.
\begin{proposition}
  \label{prop:ins-diag}
 Inserter diagrams are exactly functors from the walking co-fork $⇉→$ to $\Cat$.
\end{proposition}

Given a signature $\Sigma$, its category of \kl{models} then corresponds to the \emph{vertical inserter} of the inserter diagram $(Σ,Θ,π_1)$ associated to $\Sigma$:

\begin{definition}
  The \AP\intro{vertical inserter} of an \kl{inserter diagram} $(F,G,p)$ is the
  category whose objects are pairs $(a,f)$ where $a$ is an object of $A$ and
  $f$ is a morphism $f ∶ F a → G a$ such that $p(f)$ is the identity morphism.
\end{definition}

\subsubsection{2-Categorical Structure of Inserter Diagrams}\label{subsubsec:2-cat-struct-ins}

The interest of abstracting \kl{module signatures} by \kl{inserter diagrams}
is that \kl{inserter diagrams} have a 2-categorical structure as described below.
This will then enable us to define a 2-functor computing the category of models,
which we have abstracted as \kl{vertical inserters}.

\begin{definition}
  A \emph{morphism} between two \kl{inserter diagrams} $(F,G,p)$ and $(F',G',p')$ is
  given by three functors $(A \xrightarrow{H_A} A', B \xrightarrow{H_B} B',
  C \xrightarrow{H_C} C')$ such that $H_C ∘ p = p' ∘ H_B$ and $H_B ∘ G = G' ∘ H_A$,
  together with a natural transformation
  $F'∘ H_A \xrightarrow{h_F}  H_B∘F$ as summarised in the following diagram.
  Moreover, $p' ∘ h_F$ must the identity 2-cell on $p' ∘ H_B ∘ F = p' ∘ F' ∘ H_A $.
  \[
  \begin{tikzcd}[row sep=4em, column sep=huge]
    A \arrow[r, shift left=.75ex, "F"] \arrow[r, shift right=.75ex, swap, "G"] \arrow[d, swap, "H_A"{name=HA}]
      & B \arrow[r, "p"] \arrow[d, "H_B"{name=HB}]
      & C \arrow[d, "H_C"] \\
    A' \arrow[r, shift left=.75ex, "F'"] \arrow[r, shift right=.75ex, swap, "G'"]
      & B' \arrow[r, "p'"]
      & C'
      \arrow[from=HA, to=HB, "h_F"', swap, Rightarrow, shorten <=2em, shorten >=2em]
  \end{tikzcd}
  \]
\end{definition}

\begin{remark}
  This notion of morphism may seem ad-hoc, but it naturally appears
  when building the 2-functor computing the category of models.
  It also enables us to understand vertical inserters in terms of \emph{marked limits},
  as explained in \cref{rem:vins-as-limits} below.
\end{remark}

Let us now complete the definition of the 2-category of vertical inserter diagrams.

\begin{definition}
  We define the 2-category \AP $\intro*\VInsDiag$ of vertical inserter diagrams, where
  objects and morphisms are as above, and a 2-cell between $(H_A, H_B, H_C, h_F)$ and
  $(H_A', H_B', H_C', h_F')$ is given by three natural transformations
  $(H_A \xrightarrow{h_A}H_A', H_B \xrightarrow{h_B}H_B', H_C \xrightarrow{h_C}H_C' )$
  satisfying coherence conditions, making them modifications~\cite[4.4]{twodimcat}
  between the lax transformations induced by \cref{prop:ins-diag}.
\end{definition}

\subsubsection{Models from Inserter Diagrams, 2-Functorially}\label{subsubsec:models-from-inserter-diags}

The main interest of the 2-categorical structure of \kl{inserter diagrams} is
that \kl{vertical inserters}, abstracting models, can be computed by a 2-functor.
Consequently, adjunctions or equivalences in the 2-category $\VInsDiag$ lift to
adjunctions or equivalences between vertical inserters, that is, between categories of models.

\begin{theorem}
  \label{thm:def-vins}
  The 2-functor \AP $\intro*\CstDiag ∶ \Cat → \VInsDiag$ mapping a category $\C$ to the trivial
  diagram $\C ⇉ \C → \C$ has a right adjoint $\intro*\VIns$, computing the
  vertical inserter --- that is, the category of models --- of a given diagram.
  \[ \VInsDiag \xrightarrow{\VIns} \Cat \]
\end{theorem}

\begin{remark}
  \label{rem:vins-as-limits}
  \cref{thm:def-vins} exploits the characterisation of vertical inserters
  as 2-dimensional limits, and
  is analogous to the fact that a category $\C$ has limits of shape $\D$ precisely when
  the functor $\C → [\D,\C]$ mapping $c$ to the constant functor equal to $c$ has a right adjoint,
  which computes the limit.
  Indeed, if we take $\C = \Cat$ and $\D$ the walking cofork $⇉→$, we almost get the same situation,
  except that we need to replace the strict morphisms of $[⇉→, \Cat]$ by
  \emph{lax natural transformations}~\cite[Chapter 4.2]{twodimcat} such that the only non-strict component of the
  natural transformation is the one associated with top left arrow.
  Any other choice of "laxity" in the notion of morphism of inserter diagrams would
  not yield a right adjoint.

  This adjunction holds more generally for \emph{marked limits}, or \emph{cartesian quasi-limits} in the terminology of \citeauthor{GrayMarkedLimits} \cite[I.7.9.1.(iii)]{GrayMarkedLimits}.
\end{remark}

Given a \kl{module signature} $Σ$, there is a projection $π_1 : \Model(Σ) → \C$ mapping
\kl{models} $(R,m)$ to $R$.
This extends to a 2-natural transformation by requiring it to be compatible with
the projection for \kl{inserter diagrams} as defined below.
Naturality is important, as for instance, it ensures that the functor induced
between the categories of models is compatible with the projection to the
category of monoids.

\begin{theorem}
  There is a 2-functor $π_1 : \VInsDiag → \Cat$ mapping an inserter diagram of the form
  $\begin{tikzcd} A \doublerightarrow{r}{F}{G} & B \ar[r, "p"] & C \end{tikzcd}$
  to the category $A$.
\end{theorem}

\subsection{Models from Module Signatures, 2-Functorially}
\label{subsec:models-modules}

We have shown how the categories of \kl{models} of a \kl{module signatures}
abstract as \kl{vertical inserters} of \kl{inserter diagrams}, and that
computing models of such signatures is 2-functorial.
However, in practice, we want to lift adjunctions or equivalences directly
from monoidal categories and module signatures, not from inserter diagrams.
To make this possible, we now precompose the 2-functor $\VIns : \VInsDiag → \Cat$
by a 2-functor $\PModDiag : \ModSig → \VInsDiag$ mapping a module
signature to the associated inserter diagram.

To define a 2-category $\ModSig$ whose objects are pairs $(\C,Σ)$ of a
monoidal category $\C$ and a module signature $Σ$ over it, we first need
to show that lax monoidal functors lift to \kl{monoids} and \kl{modules}.

\begin{proposition}[{\cite[\S\S~4.5-4.6]{ZsidoPhd10}}]
  \label{prop:mon-functors-lift}
  A lax monoidal functor $F : \C → \D$ lifts to a functor between categories of
  \kl{monoids} $F : \Mon(\C) → \Mon(\D)$ and \kl{modules} $F : \Mod(\C) → \Mod(\D)$.
  Those liftings are compatible with composition of lax monoidal functors,
  identity functors, projections to monoids and the tautological module signature.
  Similarly, a monoidal transformation between lax monoidal functors induces a
  compatible natural transformation between the induced functors between monoids
  and modules.
\end{proposition}

We can now define the 2-category of module signatures.

\begin{definition}
   We define the 2-category \AP $\intro*\ModSig$ of module signatures as follows:
  \begin{itemize}
    \item an object consists of a monoidal category $C$ and a module signature $Σ ∶ \Mon(\C) → \Mod(\C)$;
    \item a 1-cell between $(\C,Σ)$ and $(\D,Σ')$ consists of a monoidal functor
          $F ∶ \C → \D$ and a natural transformation $α ∶ Σ' ∘ F → F ∘ Σ $, which
          is the identity when composed with the projection to monoids $\pi_1 : \Mod(\C) → \Mon(\C)$.
          Here, $F$ corresponds to the lifting of $F$ to monoids and modules as in \cref{prop:mon-functors-lift}.
    \item a 2-cell between $(F, α)$ and $(G, β)$ consists of a monoidal natural
          transformation $γ ∶ F → G$ such that the following diagram commutes.
    \[
    \begin{tikzcd}
      Σ' ∘ F \ar[r, "α"] \ar[d, swap, "Σ' ∘ γ"] & F ∘ Σ \ar[d, "γ ∘ Σ"]
      \\
      Σ' ∘ G \ar[r, "β", swap] & G ∘ Σ
    \end{tikzcd}
    \]
  \end{itemize}
\end{definition}

With this definition, we can now show that the assignment of the \kl{inserter diagram}
to a module signature is 2-functorial, and hence that the category of models is
2-functorial in module signatures.

\begin{theorem}
  There is a 2-functor \AP $\intro*\PModDiag ∶\ModSig → \VInsDiag$
  mapping $(\C,Σ)$ to the following \kl{inserter diagram}:
  \[
    \begin{tikzcd}
      \Mon(\C) \doublerightarrow{r}{Σ}{Θ} & \Mod(\C) \ar[r, "p"] & C
    \end{tikzcd}
  \]
\end{theorem}

\begin{theorem}
  \label{thm:2-functoriality-models}
  By composing the 2-functors introduced above, we get a 2-functor \AP $\intro*\Model
  : \ModSig → \Cat$ that computes the category of models of any module signature:
  \[ \ModSig \xrightarrow{\PModDiag} \VInsDiag \xrightarrow{\VIns} \Cat \]
\end{theorem}

Though not directly related to the main topic of this paper, this functor
can be used to give a 2-categorical perspective on the initiality theorem
\cite[Theorem 4.23]{lamiaux2025unifiedframeworkinitialsemantics}.

\begin{remark}
  The 2-functor of \cref{thm:2-functoriality-models} can be restricted to the full sub-2-category
  $\NicePSigStrength$
  of module signatures
  induced by \emph{signatures with strength} satisfying the
  hypotheses of the initiality theorem \cite[Theorem 4.23]{lamiaux2025unifiedframeworkinitialsemantics}.
  Such a signature is given by an $\omega$-cocontinuous functor specifying the operations,
  and a strength used to build the monoid structure.
  The image of any such signature by this functor is a category
  which has an initial object.
  \[ \NicePSigStrength \longrightarrow \ModSig \xrightarrow{\Model} \Cat \]
\end{remark}

\section{Binding-Friendly Monoidal Categories}
\label{sec:binding-friendly-mon-cat}

\emph{Binding signatures} can be used to specify untyped languages involving variable bindings.
We are interested in relating models of binding signatures over \emph{different}
base categories, such as $[𝔽,𝐒𝐞𝐭]$ and $[𝐒𝐞𝐭,𝐒𝐞𝐭]$.
We could do so by first relating the module signatures generated by a binding signature for each pair of base categories of our interest (e.g., exhibiting an adjunction in $\ModSig$), and then applying 2-functoriality to get the same relation on the categories of models, by \cref{prop:2-functors-preserve-relations}.

Instead, we follow another path that abstracts and factors a large amount of this work: we show that binding signatures can be ``interpreted'' in any monoidal
category with enough structure, which we call ``binding-friendly'' monoidal categories, and that this interpretation is 2-functorial.
Consequently, it suffices to prove that two binding-friendly monoidal categories are related
in order to conclude that their respective models of
binding signatures are also related.

In this way we can recover, as simple consequences and generically, results that
were previously proven in an ad-hoc manner.
Specifically, in \cref{sec:application-FSet}, we show that there is a
\kl{coreflection} between models in $[𝔽, 𝐒𝐞𝐭]$ and models in $[𝐒𝐞𝐭, 𝐒𝐞𝐭]$,
and in \cref{sec:application-equivalence-debruijn} that there is an equivalence
between suitable full subcategories of models in the De Bruijn
setting~\cite{dblmcs} (see \cref{def:debruijn-monad} below about \emph{De Bruijn
monads}) and $[𝔽,𝐒𝐞𝐭]$.

\subsection{Binding Arities in a Generic Monoidal Category}
\label{subsec:binding-arities-in-monoidal}

A binding signature specifies a syntax generated by operations, potentially binding
variables in each of their arguments.

\begin{definition}[\cite{aczel:general-church-rosser}]
  A \AP\intro{binding arity} is a list of natural numbers.
  A \intro{binding signature} is a family $([n_1,…,n_{p_o}])_{o ∈ O}$ of binding arities.
\end{definition}

Each binding arity specifies a language constructor. The length of the list is the number
of arguments, and the $i$-th element is the number of variables bound by the $i$-th argument.

\begin{example}\label{ex:b-sig-LC}
  The untyped lambda calculus with constructors $\app$ and $\abs$ can be specified
  by the \kl{binding signature} $([0,0],[1])$.
  The corresponding module signature is the signature $\Theta \times \Theta + \Theta^{(1)}$ of \cref{ex:sig-LC-cont}.
\end{example}

A binding signature $((n_1,…,n_{p_o}))_{o ∈ O}$ can be interpreted in $[𝐒𝐞𝐭, 𝐒𝐞𝐭]$
as the \kl{module signature} below~\cite{hirscho:lam}.
Each summand of the coproduct represents a constructor, and the elements of the
product representing its arguments, the module signature $Θ^{(n_i)}$
representing the binding of $n_i$ variables.
This uses the trivial signature (\cref{def:theta}), and the $n$-th \AP\intro{derivative}
of the module signature $Σ$, adding $n$ fresh variables to contexts as $Σ^{(n)}(X)(Γ) := Σ(X)(Γ+n)$.
\[ ∐_{o ∈ O} ∏_{i=1}^{p_o} Θ^{(n_i)} \]

In general, to interpret a binding signature as a module signature in a monoidal
category, one has to be able to build coproducts, finite products, the trivial
signature $Θ$ and the \kl{derivative} of a module signature.
The module signature $Θ$ already exists for all monoidal categories.
Furthermore, module signatures inherit their limits and colimits from the base
category $\C$ provided that for all $Z$ precomposition by $\_ ⊗ Z$ preserves them.
This is a consequence of the following result.

\begin{theorem}
  \label{thm:monadic-param-modules}
  Given a monoidal category $\C$, the category of module signatures
  is monadic over $[\Mon(\C),\C]$, the monad mapping $Σ$ to $Σ(-)⊗-$.
\end{theorem}
\begin{proof}
  First, we check that $Σ(-) ⊗ -$ is indeed a monad.
  The unit at $Σ$ is the natural family of morphisms
  $Σ(X) \xrightarrow{ρ}Σ(X)⊗I \xrightarrow{Σ(X)⊗e} Σ(X)⊗X$, where
  $e ∶ I → X$ is the unit of the monoid $X$.

  The multiplication at $Σ$ is the natural transformation
  $(Σ(X) ⊗ X) ⊗ X \xrightarrow{α} Σ(X)⊗(X ⊗ X)
  \xrightarrow{Σ(X)⊗m} Σ(X)⊗X$, where $m ∶ X⊗X → X$ is the multiplication
  of the monoid $X$.
  It is easy to check that the monad laws hold.

  An algebra consists of a functor $Σ ∶ \Mon(\C) → \C$ and a natural family of
  morphims $x ∶ Σ(X)⊗X → X$ and the algebra laws show that this provides a
  $X$-module structure on $Σ(X)$.
  We then get a functor from algebras to $[\Mon(\C), \Mod(\C)]$;
  this functor is an isomorphism.
\end{proof}

\begin{proposition}
  \label{prop:limits-colimits-param-modules}
  Limits in the category of module signatures are computed pointwise in $C$,
  and if $- ⊗ X$ preserves colimits of shape $D$ for all $X$,
  then colimits of shape $D$ are also computed pointwise.
\end{proposition}
\begin{proof}%
  By \cref{thm:monadic-param-modules}, the functor from module signatures to
  the category $[\Mon(\C), \C]$ is monadic, and thus creates limits by
  \cite[Theorem 5.6.5]{riehl2017category}.
  Because limits are computed pointwise in a functor category, we get the
  desired result for limits.

  As for colimits of shape $D$, the involved monad $Σ ↦ Σ(-)⊗-$ preserves them
  because colimits are computed pointwise in a functor category.
  We conclude by \cite[Theorem 5.6.5]{riehl2017category}: a monadic functor creates any
  shape of colimit that is preserved by the monad.
\end{proof}

\noindent It remains to interpret the \kl{derivative} of a module signature in a
generic monoidal category.
In the case of $[𝐒𝐞𝐭,𝐒𝐞𝐭]$, this interpretation is given by precomposition
with $-+1$, which has the following characterisation.

\begin{lemma}
  \label{lemma:power-adjoint}
  Denoting the endofunctor $-+1$ by $O ∶ 𝐒𝐞𝐭 → 𝐒𝐞𝐭$,
  the endofunctor $-∘O$ on $[𝐒𝐞𝐭, 𝐒𝐞𝐭]$ is right adjoint to $- × \Id$.
\end{lemma}
\begin{proof}
  By~\cite[Proposition 3]{DBLP:conf/cie/AltenkirchLS10}, since
  the identity endofunctor is, using their notations, $⟦ 1 \triangleleft 1 ⟧$.
\end{proof}

\noindent Indeed, as exponentiation by ${(-)}^{\Id}$ is a right adjoint to $- × \Id$,
and adjoints are unique up to isomorphism, we could define instead \kl{derivation} by
$Σ^{(1)} := {Σ(-)}^{\Id}$.
This definition generalises to any monoidal category with exponentiable unit,
that is such that $- \times I$ has a right adjoint.

\begin{theorem}
  \label{thm:exponential-parammodule}
  Let $\C$ be a monoidal category with binary products which are preserved on the left
  by the tensor, such that $I$ is exponentiable.
  If $Σ$ is a module signature, then so is $Σ(-)ᴵ$,
  with module structure $Σ(X)ᴵ ⊗ X → Σ(X)ᴵ$ defined as
  the transpose of the following morphism.
  \[
    (Σ(X)ᴵ ⊗ X) × I →
    (Σ(X)ᴵ ⊗ X) × (I ⊗ X)
    ≅
    (Σ(X)ᴵ × I) ⊗ X
    →
    Σ(X) ⊗ X
    → Σ(X)
  \]
  Moreover,
  $Σ(-)^I$ is actually the
  exponential $Σ^Θ$ in the category of \kl{module signatures}.
\end{theorem}
\begin{proof}
  See \cref{app:exponential-parammodule}.
\end{proof}

\begin{notation}
In the situation above, given a module signature $ Σ$, we denote
$Σ^Θ$ by $Σ^{(1)}$, $(Σ^Θ)^{Θ}$ by $Σ^{(2)}$, and so on.
\end{notation}

\subsection{Binding-Friendly Monoidal Categories}

The discussion in \cref{subsec:binding-arities-in-monoidal} leads us to introduce the notion of binding-friendly monoidal categories,
which contains sufficient structure to interpret binding signatures.
Such categories, and suitable morphisms and 2-cells between them, form a 2-category (\cref{prop:2-cat-binding-friendly}).

\begin{definition}
  A monoidal category $\C$ is said \AP\intro(monoidal){binding-friendly} if it has
  \begin{itemize}
    \item finite products left-preserved by the tensor;
    \item non-empty coproducts left-preserved by the tensor;
    \item an exponentiable unit $I$.
   \end{itemize}
\end{definition}

\begin{remark}
    It would be more convenient to ensure that empty coproducts exist, so that
    the empty binding signature induces a module signature like any other.
    However, we will encounter a situation in \cref{sec:application-equivalence-debruijn}
    where this is not the case (see \cref{prop:fset-bnset-equivalent} below).
\end{remark}

\begin{example}
  The category $[𝐒𝐞𝐭,𝐒𝐞𝐭]$ is \kl(monoidal){binding-friendly}, by \cref{lemma:power-adjoint}
\end{example}

\kl{Binding-friendly monoidal} categories naturally inherit a 2-categorical structure from
monoidal categories by requiring the extra structure to be preserved by the morphisms.

\begin{definition}
  \label{def:binding-friendly-functor}
  A binding-friendly functor $F$ between two binding-friendly monoidal
  categories consists of a lax monoidal functor preserving finite products,
  small coproducts, and exponentiation by the unit, in the sense that for every
  object $X$ of the domain, the canonical morphism $F(Xᴵ) → F(X)ᴵ$, obtained as
  the transpose of the below morphism, is an isomorphism.
  \[ \begin{tikzcd}
    F(Xᴵ) × I \ar[r]
    & F(Xᴵ) × FI ≅ F(Xᴵ × I) \ar[r]
    & F(X)
  \end{tikzcd} \]
\end{definition}
\begin{definition}
  A transformation between binding-friendly functors is a monoidal
  transformation between the underlying monoidal functors.
\end{definition}
\begin{proposition}\label{prop:2-cat-binding-friendly}
  This defines a 2-category \AP $\intro*\BindMonCat$ of binding-friendly
  monoidal categories, which comes equipped with a forgetful 2-functor to the 2-category
  of monoidal categories.
\end{proposition}

\subsection{Interpreting Binding Signatures}

The point of binding-friendly monoidal categories is that there is a
canonical 2-functorial interpretation of binding signatures into them.
Here, the existence of products and coproducts in the binding-friendly monoidal category
ensures the existence of products and coproducts of modules, and the existence of an exponentiable
unit ensures the existence of the derivative module $Θ^{(n)}$ representing the binding
of $n$-variables, by \cref{thm:exponential-parammodule}.

\begin{definition}
  \label{def:bindsig-module}
  Given any non-empty binding signature $S=((n_1,…,n_{p_o}))_{o ∈ O}$ and a
  binding-friendly monoidal category $C$, we define the module signature $S_C$ as
  \[ ∐_{o ∈ O} ∏_{i=1}^{p_o} Θ^{(n_i)} \]
\end{definition}

\noindent Moreover, for any non-empty binding signature, this mapping from binding-friendly monoidal categories to module signatures extends to a 2-functor. By post-composing it with the 2-functor $\Model$, we then get the category of models in any given binding-friendly monoidal category, 2-functorially.

\begin{proposition}
  \label{prop:param-module-funct}
  Any non-empty binding signature $S$ induces a 2-functor $\intro*\Sem{S} ∶
  \BindMonCat → \ModSig$ mapping a binding-friendly monoidal category $C$ to
  the module signature $S_C$.
\end{proposition}
\begin{proof}
  See \cref{app:bindingfriendly-sig}.
\end{proof}

\begin{theorem}
  \label{thm:2-functoriality-bindingfriendly}
  Any binding signature $S$ induces a 2-functor \AP$\intro*\BSigModel{S} ∶ \BindMonCat → \Cat$ mapping a
  binding-friendly monoidal category to its category of models, defined in the
  non empty case by the composition:
  \[
    \begin{tikzcd}[label={eq:2-functoriality-bindingfriendly}, column sep=large]
      \BindMonCat \ar[r, "\Sem{S}"]
      & \ModSig \ar[r, "\Model", scale=4]
      & \Cat
    \end{tikzcd}
  \]
\end{theorem}
\begin{proof}
  2-functoriality for the empty signature is trivial as its category of model is just the category of monoids.
  For the non-empty case, it follows by composition.
\end{proof}

\section{Application: Relating Models in \texorpdfstring{$[𝐒𝐞𝐭, 𝐒𝐞𝐭]$}{[Set,Set]} and \texorpdfstring{$[𝔽, 𝐒𝐞𝐭]$}{[F, Set]}}
\label{sec:application-FSet}

In her PhD dissertation \cite[Chapter 4.11]{ZsidoPhd10}, Zsidó related initial
semantics of \kl{binding signatures} in $[𝐒𝐞𝐭, 𝐒𝐞𝐭]$ and in the monoidal
category $[𝔽, 𝐒𝐞𝐭]$ introduced by Fiore et al.~\cite{DBLP:conf/lics/FiorePT99}.
More specifically, she lifted the monoidal adjunction between $[𝔽, 𝐒𝐞𝐭]$ and
$[𝐒𝐞𝐭, 𝐒𝐞𝐭]$ to monoids and uses it directly to build the initial model
of $[\F,\Set]$ out of the one on $[\Set,\Set]$, and vice versa.

We generalize this result, and provide a direct proof of it, using 2-functoriality.
We lift the monoidal adjunction $[𝔽,𝐒𝐞𝐭] \simeq [𝐒𝐞𝐭,𝐒𝐞𝐭]_f \leftrightarrows [𝐒𝐞𝐭,𝐒𝐞𝐭]$
to the 2-category $\BindMonCat$, and hence obtain an adjunction between the
respective categories of models by 2-functoriality (\cref{prop:2-functors-preserve-relations}).
We then recover Zsidó results as a corollary, since this adjunction is actually
a \kl{coreflection}, and therefore preserves initial objects.
In summary, we construct the following functors:
\[ \BSigModel{S}([𝔽, 𝐒𝐞𝐭]) ≃ \BSigModel{S}([𝐒𝐞𝐭,𝐒𝐞𝐭]_f) \leftrightarrows \BSigModel{S}([𝐒𝐞𝐭,𝐒𝐞𝐭]).\]

\subsection{The Coreflection Lifts to \texorpdfstring{$\BindMonCat$}{BindMonCat}}

Let us first review the adjunction between $[𝔽, 𝐒𝐞𝐭]$ and
$[\Set,\Set]$ in the 2-category of categories.

\begin{proposition}
  \label{prop:coreflection-FSet}
  Let $J : 𝔽 → 𝐒𝐞𝐭$ be the canonical embedding. There is a
  \kl{coreflection} $[𝔽, 𝐒𝐞𝐭] \stackrel{- ∘ J}{\leftrightarrows}
  [𝐒𝐞𝐭,𝐒𝐞𝐭]$ which factorises as an equivalence of categories followed by
  a coreflective embedding
  \begin{equation}
    \label{eq:decomposition-coreflection-FSet}
    [𝔽, 𝐒𝐞𝐭] \simeq [𝐒𝐞𝐭,𝐒𝐞𝐭]_f \leftrightarrows [𝐒𝐞𝐭,𝐒𝐞𝐭],
  \end{equation}
  where the left adjoint to precomposition is the left Kan extension, and $[𝐒𝐞𝐭, 𝐒𝐞𝐭]_f$
  denotes the full subcategory of \emph{finitary endofunctors} of $𝐒𝐞𝐭$, that is,
  endofunctors that preserve \emph{filtered colimits}~\cite[Definition 1.4]{Adamek}.
\end{proposition}
\begin{proof}
  Since $J$ is full and faithful, the left Kan extension $\Lan_J$ is also full and faithful
  by \cite[Corollary 6.3.9 and Lemma 4.5.13]{riehl2017category}.
  Thus $\Lan_J$ factors as an equivalence with its image, which is $[𝐒𝐞𝐭,
  𝐒𝐞𝐭]_f$, and its \kl{coreflection}. For instance, see \cite[Proposition 4.2]{dblmcs}.
\end{proof}

To lift the adjunction to $\BindMonCat$, we must first show that the different
categories involved are \kl{binding-friendly monoidal} categories.

\begin{proposition}
  \label{prop:finitary-submonoidal}
  $[𝐒𝐞𝐭, 𝐒𝐞𝐭]_f$ is a \kl{binding-friendly monoidal} subcategory of $[𝐒𝐞𝐭, 𝐒𝐞𝐭]$.
\end{proposition}
\begin{proof}
   Finitary endofunctors are closed under composition, colimits, finite
  products, and exponentials by $\Id$ (which is precomposition by $-+1$).
\end{proof}

By the equivalence between $[𝔽, 𝐒𝐞𝐭]$ and $[𝐒𝐞𝐭,𝐒𝐞𝐭]_f$, we get a
\kl{binding-friendly monoidal} structure on $[𝔽, 𝐒𝐞𝐭]$, which is used
by \citet{DBLP:conf/lics/FiorePT99} to define their category of models
of a binding signature.

\begin{corollary}
  The category $[𝔽, 𝐒𝐞𝐭]$ is \kl{binding-friendly monoidal}
  with $-ᴵ$ given by precomposition by $- + 1$.
\end{corollary}
\begin{proof}
  To check what $-ᴵ$ does, we consider $X ∶ 𝔽 → 𝐒𝐞𝐭$, apply the left Kan
  extension, precompose with $-+1$, which is $-ᴵ$ in $[𝐒𝐞𝐭, 𝐒𝐞𝐭]_{f}$, and
  investigate the restriction along $J$: we get what is claimed.
\end{proof}

We are now ready to lift the coreflection to $\BindMonCat$.

\begin{proposition}
  \label{prop:corefl-finitary-extends}
  The factorisation $[𝔽,𝐒𝐞𝐭] \simeq [𝐒𝐞𝐭,𝐒𝐞𝐭]_f \leftrightarrows [𝐒𝐞𝐭,𝐒𝐞𝐭]$ as an
  equivalence followed by a \kl{coreflection} lifts in the 2-category $\BindMonCat$.
\end{proposition}
\begin{proof}
  The equivalence in $\BindMonCat$ between $[𝔽, 𝐒𝐞𝐭]$ and $[𝐒𝐞𝐭,𝐒𝐞𝐭]_f$ comes
  from the fact that the \kl{binding-friendly monoidal} structure on $[𝔽, 𝐒𝐞𝐭]$
  is defined by transporting the one of $[𝐒𝐞𝐭, 𝐒𝐞𝐭]_f$ along the equivalence.

  \cref{prop:finitary-submonoidal} shows that the embedding $[𝐒𝐞𝐭, 𝐒𝐞𝐭]_f ↪ [𝐒𝐞𝐭, 𝐒𝐞𝐭]$
  lifts in $\BindMonCat$, so what remains to show is that the right adjoint
  $[𝐒𝐞𝐭, 𝐒𝐞𝐭] \xrightarrow{- ∘ J} [𝔽, 𝐒𝐞𝐭] ≃ [𝐒𝐞𝐭, 𝐒𝐞𝐭]_f$ is \kl(monoidal){binding-friendly}.
  Preservations of finite products and coproducts follows from precomposition
  being continuous and cocontinuous; lax monoidality follows from being right
  adjoint to an oplax monoidal functor.
  Finally, exponentiation by the unit is preserved by the right adjoint.
  This is easy to check since in all the involved categories, this exponentiation is given
  by precomposition with $- + 1$.
\end{proof}

\subsection{The Coreflection between the Categories of Models}

Using 2-functoriality of $\BSigModel{S}$ (\cref{thm:2-functoriality-bindingfriendly}),
we can now directly lift the adjunction to the categories of models.

\begin{theorem}
  \label{thm:lifting-models}
  The factorisation of \cref{eq:decomposition-coreflection-FSet} lifts to the category of models, so that for any
  binding signature $S$, there is an equivalence and a \kl{coreflection} as follows:
  \begin{equation}
    \label{eq:lifting-models}
    \BSigModel{S}([𝔽, 𝐒𝐞𝐭]) ≃ \BSigModel{S}([𝐒𝐞𝐭,𝐒𝐞𝐭]_f) \leftrightarrows \BSigModel{S}([𝐒𝐞𝐭,𝐒𝐞𝐭]).
  \end{equation}
\end{theorem}
\begin{proof}
  As a 2-functor, $\BSigModel{S}$ preserves equivalences and \kl{coreflection}s.
\end{proof}

\noindent We can now recover Zsidó's result as coreflections preserve initial objects.

\begin{corollary}
  \label{cor:zsido-preserves-initial}
  All the back-and-forth functors in \cref{eq:lifting-models} preserve initial objects.
\end{corollary}
\begin{proof}
  Left adjoints and equivalences preserve colimits: this proves the case of all
  the above functors except $\BSigModel{S}([𝐒𝐞𝐭,𝐒𝐞𝐭]) → \BSigModel{S}([𝐒𝐞𝐭,𝐒𝐞𝐭]_f)$.
  But the right adjoint $R$ of a \kl{coreflection} $L$ also preserves the
  initial object because then the unit component $0 → RL0$ is an isomorphism~\cite[Lemma 4.5.13]{riehl2017category}.
\end{proof}

\section{Application: Relating Models in \texorpdfstring{$[𝔽, 𝐒𝐞𝐭]$}{[F, Set]} and the De Bruijn Setting}
\label{sec:application-equivalence-debruijn}

In this section, we recover a new proof of the restricted equivalence between
\emph{De Bruijn models}~\cite{dblmcs} and models in $[𝔽, 𝐒𝐞𝐭]$.
The two notions of models are exemplified in \cref{code:ULC-unscoped,code:ULC-wscoped-finite} for the lambda calculus.
The crucial difference is that a De Bruijn model does not provide explicit
information about the support of terms, while in $[𝔽, 𝐒𝐞𝐭]$, terms come with
their scope of available free variables.

Nonetheless, given any binding signature, De Bruijn models are
equivalent to models in $[𝔽, 𝐒𝐞𝐭]$, provided that both are restricted to an
appropriate notion of well-behavedness.
Intuitively, these conditions amount, on the De Bruijn side, to restricting De Bruijn models to using only finitely many variables.
On the finite-context side, we restrict to models that map the empty context to the set of closed terms
(see \cref{ex:non-intersectional} below for a counter-example).

Once these definitions are in place, we can prove the following result:

\begin{theorem*}[{\cite[Theorem 4.25]{dblmcs}}]
  \label{thm:lifting-models-debruijn-informal}
  Given any binding signature, the full subcategory of well-behaved
  models in $[\BN, 𝐒𝐞𝐭]$ is equivalent to the full subcategory of
  well-behaved models in $[𝔽, 𝐒𝐞𝐭]$.
\end{theorem*}

We state and prove this result formally below, in \cref{thm:lifting-models-debruijn},
using the 2-functorial theory we have developed.

\subsection{The De Bruijn Setting}

The framework of De Bruijn monads as developed by \citet{dblmcs} relies on a
\emph{skew-monoidal} category, yet our 2-categorical framework 
involve monoidal categories only. To bridge the gap, 
we show in this section how their skew-monoidal category
can be replaced by a binding-friendly monoidal category.

\subsubsection{The Binding-Friendly Monoidal Category $[\BN, 𝐒𝐞𝐭]$}

To model De Bruijn monads, we define the monoidal category $[\BN, 𝐒𝐞𝐭]$ using
left Kan extension, similarly to $[𝔽, 𝐒𝐞𝐭]$, but using instead the full
subcategory $\BN$ of $𝐒𝐞𝐭$ that has $ℕ$ as its single object.

\begin{proposition}
  Let \AP $\intro*\BN$ be the full subcategory of sets with $ℕ$ as its single object.
  Denoting the canonical embedding $\BN → 𝐒𝐞𝐭$ by $J$, there is a \kl{coreflection}
  $[\BN, 𝐒𝐞𝐭] \stackrel{- ∘ J}{\leftrightarrows} [𝐒𝐞𝐭,𝐒𝐞𝐭]$ which factorises as an
  equivalence of categories followed by a coreflective embedding
  \begin{equation}
    [\BN, 𝐒𝐞𝐭] \simeq [𝐒𝐞𝐭,𝐒𝐞𝐭]_{ω^+, cc_0} \leftrightarrows [𝐒𝐞𝐭,𝐒𝐞𝐭],
  \end{equation}
  where
  \begin{itemize}
    \item the left adjoint to precomposition is the left Kan extension;
    \item $[𝐒𝐞𝐭, 𝐒𝐞𝐭]_{ω⁺,cc_0}$ denotes the full subcategory of
          endofunctors on $𝐒𝐞𝐭$ preserving the empty set as well as colimits
          of \emph{$ω⁺$-filtered colimits}~\cite[1.21]{Adamek}, for $ω⁺$ the
          successor cardinal of $ω$.
  \end{itemize}
\end{proposition}
\begin{proof}
  The proof is similar to that of \cref{prop:coreflection-FSet}.
  The essential image of the left Kan extension was shown to be
  $[𝐒𝐞𝐭, 𝐒𝐞𝐭]_{ω^+, cc_0 }$ in the proof by \citet[Lemma 4.29]{dblmcs}.
\end{proof}

\begin{proposition}
  \label{prop:omega-plus-submonoidal}
  $[𝐒𝐞𝐭, 𝐒𝐞𝐭]_{ω⁺,cc_0}$ is a \kl{binding-friendly monoidal} subcategory of $[𝐒𝐞𝐭, 𝐒𝐞𝐭]$.
\end{proposition}
\begin{proof}
   This subcategory is closed under composition, colimits, finite products, and
  exponentials by $\Id$ (which is precomposition by $-+1$).
\end{proof}

\noindent By the equivalence between $[\BN, 𝐒𝐞𝐭]$ and $[𝐒𝐞𝐭,𝐒𝐞𝐭]_{ω⁺,cc_0}$, we get
a \kl{binding-friendly monoidal} structure on $[\BN, 𝐒𝐞𝐭]$, which we can use
to define the category of models of a binding signature.

\begin{corollary}
  The category $[\BN, 𝐒𝐞𝐭]$ is \kl{binding-friendly monoidal}
  with $-ᴵ$ given by precomposition with $suc ∶ \BN → \BN$,
  where $suc(f)$ is $f(- + 1)$ for any $f ∶ ℕ → ℕ$.
\end{corollary}
\begin{proof}
  To check what $-ᴵ$ does, we consider $X ∶ \BN → 𝐒𝐞𝐭$, apply the left Kan
  extension, precompose with $-+1$, which is $-ᴵ$ in $[𝐒𝐞𝐭, 𝐒𝐞𝐭]_{ω^+,cc_0 }$,
  and investigate the restriction at $ℕ$: we get what is claimed.
\end{proof}

\noindent Similarly to \cref{thm:lifting-models} for $[𝔽, 𝐒𝐞𝐭]$, we then obtain, by 2-functoriality
of $\BSigModel{S}$, that the equivalence and the coreflection lift to models.

\begin{proposition}
  \label{prop:corefl-omage-extends}
  The factorisation $[\BN,𝐒𝐞𝐭] \simeq [𝐒𝐞𝐭,𝐒𝐞𝐭]_{ω^+,cc_0} \leftrightarrows [𝐒𝐞𝐭,𝐒𝐞𝐭]$ as an
  equivalence followed by a \kl{coreflection} lifts in the 2-category $\BindMonCat$.
\end{proposition}
\begin{proof}
  C.f.\, \cref{prop:corefl-finitary-extends}.
\end{proof}

\begin{theorem}
  For any binding signature $S$, there is an equivalence and a \kl{coreflection} as follows:
  \[ \BSigModel{S}([𝐒𝐞𝐭,𝐒𝐞𝐭]) \leftrightarrows \BSigModel{S}([𝐒𝐞𝐭, 𝐒𝐞𝐭]_{ω^+,cc_0}) ≃ \BSigModel{S}([[\BN, 𝐒𝐞𝐭]]) \]
\end{theorem}

\subsubsection{De Bruijn Monads, and $[\BN, 𝐒𝐞𝐭]$}

The De Bruijn Monads used by \citet{dblmcs} to model \cref{code:ULC-unscoped} are
by definition relative monads for the inclusion $\BN ↪ 𝐒𝐞𝐭$.
As a consequence (see \cref{prop:debruijn-monads-monoids} below), they are equivalent to monoids in $[\BN, 𝐒𝐞𝐭]$.

\begin{definition}
  \label{def:debruijn-monad}
  A De Bruijn monad consists of a set $X$ (the image of $ℕ$) equipped with a
  \emph{simultaneous substitution operation} $\_ [ \_ ] : X × X^ℕ → X$ and a
  \emph{variable function} $η ∶ ℕ → X$.
  It must satisfy three equations: associativity $x[f][g] = x[f[g]]$, left unitality
  $x[η] = x$, and right unitality $η(n)[f][η] = x[f]$.
  A morphism between De Bruijn monads $X$ and $Y$ is a function $X → Y$ compatible with
  variables and substitution.
\end{definition}

\noindent De Bruijn monads can model syntax with variable binding in the de
Bruijn style, where variables are mere natural numbers.

\begin{proposition}
  \label{prop:debruijn-monads-monoids}
  The category of De Bruijn monads is isomorphic to the category of monoids in $[\BN, 𝐒𝐞𝐭]$.
\end{proposition}
\begin{proof}
  By definition, a De Bruijn monad is a \emph{relative monad}~\cite[Definition 2.1]{DBLP:journals/corr/AltenkirchCU14}
  on the embedding $\BN ↪ 𝐒𝐞𝐭$.
  The desired isomorphism is constructed by \citet[Theorem 3.5]{DBLP:journals/corr/AltenkirchCU14}.
\end{proof}

A similar isomorphism holds for De Bruijn modules and modules over the
corresponding monoid in $[\BN, 𝐒𝐞𝐭]$, see \cref{app:debruijn-modules}.
This justifies that our categories of models induced by the \kl{binding-friendly monoidal}
structure coincide with them (up to isomorphism).

\subsection{The Restricted Equivalence}

By \cref{prop:corefl-finitary-extends,prop:corefl-omage-extends}, we have the
coreflection in $[𝔽, 𝐒𝐞𝐭] ≃ ([𝐒𝐞𝐭,𝐒𝐞𝐭]_f) \leftrightarrows [𝐒𝐞𝐭,𝐒𝐞𝐭]$
and $[\BN, 𝐒𝐞𝐭] ≃ ([𝐒𝐞𝐭,𝐒𝐞𝐭]_{ω^+,cc_0 }) \leftrightarrows [\BN,𝐒𝐞𝐭]$
in $\BindMonCat$.
Consequently, by 2-functoriality, the difference between using $[𝔽, 𝐒𝐞𝐭]$
and $[\BN, 𝐒𝐞𝐭]$ to model syntax boils down to the difference between
$[𝐒𝐞𝐭,𝐒𝐞𝐭]_f$ and $[𝐒𝐞𝐭,𝐒𝐞𝐭]_{ω^+,cc_0}$ as binding-friendly monoidal categories.
These categories are barely different, and therefore, models in $[\BN, 𝐒𝐞𝐭]$ and $[𝔽, 𝐒𝐞𝐭]$
only differ in ill-behaved models, which we need to exclude to establish an equivalence.

\subsubsection{Restricting $[\BN, 𝐒𝐞𝐭]$ and $[𝔽, 𝐒𝐞𝐭]$}

Since objects of $[𝔽, 𝐒𝐞𝐭]$ correspond to finitary endofunctors on sets, it is natural to consider a finitary restriction on objects $[\BN, 𝐒𝐞𝐭]$.

\begin{definition}
  A functor $X ∶ \BN → 𝐒𝐞𝐭$ is said \AP\intro(debruijn){finitary} if every
  element $t ∈ X(ℕ)$ has a \intro{finite support}, that is, there exists $n ∈ ℕ$
  such that for any $f ∶ ℕ → ℕ$, if $f(i) = i$ for all $i < n$, then $X(f)(t) = t$.
\end{definition}

\begin{example}
  Let us give examples of non-finitary functors.
  Consider the De Bruijn encoding of infinitary $λ$-terms, that is, we allow infinitely deep syntax trees.
  Let us denote the embedding of variables $ ℕ → X$ by $var$.
  This is not finitary since the term $app(var(1), app(var(2), …))$ does not have a finite support.

  If we consider a syntax with an infinitary operation $op ∶ X^ℕ → X$ and
  variables $var ∶ℕ → X$, then the corresponding functor $\BN → 𝐒𝐞𝐭$ is not
  finitary, since $op(var(1), var(2),…) ∈ X$ does not have a finite support.
\end{example}

To restrict $[𝔽,𝐒𝐞𝐭]$, we need to exclude functors that are ill-behaved on the empty set.

\begin{definition}[{\cite[Definition 4.15]{dblmcs}}]
  A functor $F : 𝔽 → D$ or $F : 𝐒𝐞𝐭 → D$ is called \AP\intro{intersectional}
  if it preserves the following equaliser.
  \[
  \begin{tikzcd}
  0 \ar[r] & 1 \doublerightarrow{r}{0}{1} & 2
  \end{tikzcd}
  \]
\end{definition}

\noindent Intersectional functors behave nicely with respect to evaluation at $0$.

\begin{proposition}[{\cite[Lemma 4.14]{dblmcs}}]
  A functor $F ∶ 𝔽 → 𝐒𝐞𝐭$ is intersectional if it preserves empty intersections.
\end{proposition}

\begin{example}
  \label{ex:non-intersectional}
  Let us give an example of a non-intersectional functor.
  Consider the functor $F ∶ 𝔽 → 𝐒𝐞𝐭$ mapping $0$ to $∅$ and $n$ to the set
  of λ-terms taking free variables in $\{1,…,n\}$.
  This induces a relative monad, but it is not a model of the binding signature
  of $λ$-calculus since it no longer has an abstraction operation $F(-+1) → F$
  --- indeed, what would be its $0$-component?
  However, it still has an application operation $F × F → F$ and thus
  is a model of the binding signature specifying a binary operation.
\end{example}

\begin{remark}
  Any model in $[𝔽, 𝐒𝐞𝐭]$ is intersectional as soon as it has at least one closed term~\cite[Remark 4.19]{dblmcs}.
  This is the case in particular if there is a binding operation $(1)$, or a constant $()$.
\end{remark}

\subsubsection{Recovering The Equivalence The Between Categories of Models}
Using 2-functoriality, the key resut to prove the equivalence between the (restricted) categories of models is that the restricted subcategories  of $[𝐁ℕ, 𝐒𝐞𝐭]$ and $[𝔽, 𝐒𝐞𝐭]$ are monoidally equivalent. Of course, this first requires to show that they are binding-friendly, in particular, that they are
stable by monoidal product.

\begin{proposition}
  \label{prop:fset-bnset-equivalent}
  The full subcategory of non-empty \kl(debruijn){finitary} functors $\BN →
  𝐒𝐞𝐭$ is equivalent in $\BindMonCat$ to the full subcategory of non-empty
  \kl{intersectional} functors from $𝔽 → 𝐒𝐞𝐭$.
\end{proposition}
\begin{proof}
  C.f. \cref{subsec:proof-fset-bnset-equivalent}.
\end{proof}

\noindent We then recover a simple proof of the main result:

\begin{theorem}[{\cite[Theorem 4.25]{dblmcs}}]
  \label{thm:lifting-models-debruijn}
  Given any binding signature, the full subcategory of \kl(debruijn){finitary}
  models in $[\BN, 𝐒𝐞𝐭]$ is equivalent to the full subcategory of
  \kl{intersectional} models in $[𝔽, 𝐒𝐞𝐭]$.
\end{theorem}
\begin{proof}
  By application of the 2-functor from \cref{thm:2-functoriality-bindingfriendly}
  to the equivalence of \cref{prop:fset-bnset-equivalent}, we get an equivalence
  between non-empty \kl(debruijn){finitary} models in $[\BN, 𝐒𝐞𝐭]$ and non-empty
  \kl{intersectional} models in $[𝔽, 𝐒𝐞𝐭]$.
  But since the initial functor is not a relative monad anyway, the restriction
  to non-empty models is actually unnecessary.
\end{proof}

\noindent This equivalence maps a De Bruijn model $X$ to the functor $𝔽 → 𝐒𝐞𝐭$ mapping
$n$ to the subset of elements with support $n$; in the other direction it
intuitively maps a functor $F ∶ 𝔽 → 𝐒𝐞𝐭$ to the union of all the sets $F(n)$
(see \cref{lem:FtoBN} below for a more formal statement).

\subsection{Proof of \texorpdfstring{\cref{prop:fset-bnset-equivalent}}{Proposition \ref{prop:fset-bnset-equivalent}}}
\label{subsec:proof-fset-bnset-equivalent}

First, we have an adjunction between $[\BN, 𝐒𝐞𝐭]$ and $[𝔽,𝐒𝐞𝐭]$ defined as follows,
where $*$ denotes precomposition by inclusion; and $\Lan$ and $\Ran$ denote
Kan extensions along inclusion.
\begin{equation}
  \label{eq:adjunction-BN-F}
  \begin{tikzpicture}
\draw[white,-,curve={ratio=-0.2}, line width=0.20833333333333334em] (16.536458333333336em,-7.708333333333334em) .. controls (18.880208333333332em,-6.770833333333334em) and (21.223958333333332em,-6.770833333333334em) .. (23.567708333333336em,-7.708333333333334em);
\draw[black,->, curve={ratio=-0.2}, ] (16.536458333333336em,-7.708333333333334em) .. controls (18.880208333333332em,-6.770833333333334em) and (21.223958333333332em,-6.770833333333334em) .. (23.567708333333336em,-7.708333333333334em);
\draw[white,-,curve={ratio=-0.2}, line width=0.20833333333333334em] (23.567708333333336em,-9.479166666666668em) .. controls (21.223958333333332em,-10.416666666666666em) and (18.880208333333332em,-10.416666666666666em) .. (16.536458333333336em,-9.479166666666668em);
\draw[black,->, curve={ratio=-0.2}, ] (23.567708333333336em,-9.479166666666668em) .. controls (21.223958333333332em,-10.416666666666666em) and (18.880208333333332em,-10.416666666666666em) .. (16.536458333333336em,-9.479166666666668em);
\draw[white,-,curve={ratio=-0.2}, line width=0.20833333333333334em] (27.994791666666668em,-7.708333333333334em) .. controls (30.469444444444445em,-6.770833333333334em) and (32.92604166666666em,-6.815972222222223em) .. (35.364583333333336em,-7.84375em);
\draw[black,->, curve={ratio=-0.2}, ] (27.994791666666668em,-7.708333333333334em) .. controls (30.469444444444445em,-6.770833333333334em) and (32.92604166666666em,-6.815972222222223em) .. (35.364583333333336em,-7.84375em);
\draw[white,-,curve={ratio=-0.2}, line width=0.20833333333333334em] (35.364583333333336em,-9.34375em) .. controls (32.92604166666666em,-10.371527777777779em) and (30.469444444444445em,-10.416666666666666em) .. (27.994791666666668em,-9.479166666666668em);
\draw[black,->, curve={ratio=-0.2}, ] (35.364583333333336em,-9.34375em) .. controls (32.92604166666666em,-10.371527777777779em) and (30.469444444444445em,-10.416666666666666em) .. (27.994791666666668em,-9.479166666666668em);
\draw[white,draw=none, line width=0.20833333333333334em] (20.052083333333336em,-7.526041666666667em) .. controls (20.052083333333332em,-8.237847222222221em) and (20.052083333333332em,-8.949652777777777em) .. (20.052083333333336em,-9.661458333333334em);
\draw[black,draw=none, ] (20.052083333333336em,-7.526041666666667em) .. controls (20.052083333333332em,-8.237847222222221em) and (20.052083333333332em,-8.949652777777777em) .. (20.052083333333336em,-9.661458333333334em);
\draw[white,draw=none, line width=0.20833333333333334em] (31.69322916666667em,-7.559895833333334em) .. controls (31.69322916666667em,-8.249131944444445em) and (31.69322916666667em,-8.938368055555557em) .. (31.69322916666667em,-9.627604166666666em);
\draw[black,draw=none, ] (31.69322916666667em,-7.559895833333334em) .. controls (31.69322916666667em,-8.249131944444445em) and (31.69322916666667em,-8.938368055555557em) .. (31.69322916666667em,-9.627604166666666em);
\node at (14.322916666666668em,-8.59375em) {$[\BN, 𝐒𝐞𝐭]$} ;
\node at (25.78125em,-8.59375em) {$[𝐒𝐞𝐭, 𝐒𝐞𝐭]$} ;
\node at (37.239583333333336em,-8.59375em) {$[𝔽, 𝐒𝐞𝐭]$} ;
\node[scale=0.7] at (20.052083333333336em,-6.533676654218336em) {$\Ran$} ;
\node[scale=0.7] at (20.052083333333336em,-10.69902522448835em) {$*$} ;
\node[scale=0.7] at (31.702471338189007em,-6.536075088303378em) {$*$} ;
\node[scale=0.7] at (31.70231307124025em,-10.642811537371085em) {$\Lan$} ;
\node[rotate=-90,scale=0.7] at (20.052083333333336em,-8.59375em) {$\vdash$} ;
\node[rotate=-90,scale=0.7] at (31.69322916666667em,-8.59375em) {$\vdash$} ;
\end{tikzpicture}
\end{equation}

As always, this adjunction restricts to an equivalence between the full subcategories
of fixed points, i.e., of objects for which the (co)unit component is an isomorphism.
It turns out that the fixed points of this adjunction are precisely the finitary
functors in $[\BN, 𝐒𝐞𝐭]$ and the \kl{intersectional} functors in $[𝔽,𝐒𝐞𝐭]$, so that
we get an equivalence $[\BN, 𝐒𝐞𝐭]_{f} ≃ [𝔽,𝐒𝐞𝐭]_{int_0}$ between the two full subcategories.
Moreover, the \kl{binding-friendly monoidal} structures restrict to them, and it
is easy to check that models correspond to finitary models in $[\BN, 𝐒𝐞𝐭]$
and \kl{intersectional} models in $[𝔽, 𝐒𝐞𝐭]$.
Unfortunately, the above equivalence is not monoidal: $H ∶ [𝔽, 𝐒𝐞𝐭] → [\BN,𝐒𝐞𝐭]$
is not strong monoidal, although it is lax.
In fact, $H$ is almost strong monoidal: $H(I) →I$ is an isomorphism,
and $H(A ⊗ B) → H(A) ⊗ H(B)$ is always an isomorphism, except for $B = 0$.
This observation motivates the exclusion of the initial objects mentioned in
the statement of \cref{prop:fset-bnset-equivalent}.
Doing so, we get \kl{binding-friendly monoidal} subcategories $[\BN, 𝐒𝐞𝐭]^*_{f}$
and $[𝔽, 𝐒𝐞𝐭]^*_{int_0}$, and the above equivalence restricts to a monoidal equivalence as desired.

The rest of this section is devoted to detailing the above arguments.

\begin{lemma}
  \label{lem:FtoBN}
  The left adjoint of \cref{eq:adjunction-BN-F} maps $F ∶ 𝔽 → 𝐒𝐞𝐭$ to the
  functor $\BN → 𝐒𝐞𝐭$ mapping $ℕ$ to the colimit of the following chain.
  \[ F0 → F1 → F2 → … \]
\end{lemma}

\begin{lemma}
  Consider the assignment $X ∶ \BN → 𝐒𝐞𝐭$ to $X'∶𝔽 → 𝐒𝐞𝐭$, defined as mapping $n$
  to the subset of elements $t$ of $X(ℕ)$ with support $n$, that is, such that
  for any $f ∶ ℕ → ℕ$, if $f(i) = i$ for all $i < n$, then $X(f)(t) = t$.
  This extends to a functor which is right adjoint to $[𝔽,𝐒𝐞𝐭] → [\BN,𝐒𝐞𝐭]$.

  The counit evaluated at $X ∶ \BN → 𝐒𝐞𝐭$ is the inclusion of the elements of $X(ℕ)$ with \kl{finite support}.
  The unit evaluated at $F ∶ 𝔽 → 𝐒𝐞𝐭$ is a natural transformation which is
  the identity at each natural number $n$, except for $n = 0$ where it is the
  inclusion of $F0$ into the equaliser of $F1 ⇉ F2$.
\end{lemma}

\begin{corollary}
  The fixed points of the adjunction between $[\BN, 𝐒𝐞𝐭]$ and $[𝔽, 𝐒𝐞𝐭]$
  are the \kl(debruijn){finitary} functors in $[\BN, 𝐒𝐞𝐭]$ and the
  \kl{intersectional} functors in $[𝔽, 𝐒𝐞𝐭]$, which are thus equivalent
  subcategories.
\end{corollary}

\begin{notation}
  Given a category $C$, we denote the full subcategory of non-initial objects of
  $C$ by $C^*$, and by $[C, C]_{f}$ the full subcategory of finitary endofunctors on $C$.
\end{notation}

\begin{proposition}
  \label{prop:exclude-initial}
  $[\BN, 𝐒𝐞𝐭]^*$ (resp. $[𝔽, 𝐒𝐞𝐭]^*$) is a \kl{binding-friendly monoidal}
  subcategory of $[\BN, 𝐒𝐞𝐭]$ (resp. $[𝔽, 𝐒𝐞𝐭]$).
\end{proposition}

The following result completes the last step in the proof
of \cref{prop:fset-bnset-equivalent}.

\begin{lemma}
  The following lax monoidal functor is strong monoidal.
  \[ [𝔽, 𝐒𝐞𝐭]^* ↪ [𝔽,𝐒𝐞𝐭] \xrightarrow{\Lan} [𝐒𝐞𝐭, 𝐒𝐞𝐭] \xrightarrow{*} [\BN,𝐒𝐞𝐭] \]
\end{lemma}
\begin{proof}
  The composition of the first two functors factors as the following composition
  of strong monoidal functors, where $[𝐒𝐞𝐭, 𝐒𝐞𝐭]_{ω⁺}^*$ denotes the full
  subcategory of non-empty endofunctors preserving $ω⁺$-filtered colimits.
  \[ [𝔽, 𝐒𝐞𝐭]^* ≃ [𝐒𝐞𝐭, 𝐒𝐞𝐭]_f^* ↪ [𝐒𝐞𝐭, 𝐒𝐞𝐭]_{ω⁺}^* ↪ [𝐒𝐞𝐭,𝐒𝐞𝐭] \]
  We are left with showing that $[𝐒𝐞𝐭, 𝐒𝐞𝐭]_{ω⁺}^* ↪ [𝐒𝐞𝐭,𝐒𝐞𝐭] → [\BN,𝐒𝐞𝐭]$ is strong monoidal.
  Let us denote by $R$ the right adjoint $[𝐒𝐞𝐭, 𝐒𝐞𝐭] →[\BN,𝐒𝐞𝐭]$,
  and $L$ its left adjoint (the Kan extension).
  The lax monoidal structure on $R$ is $RA ⊗ RB = L R A ∘ RB \xrightarrow{ε_A ∘ B} A ∘ RB$.
  We want to show that when $A$ and $B$ are in $[𝐒𝐞𝐭, 𝐒𝐞𝐭]_{ω⁺}^*$, this is an isomorphism.
  It is enough to show that $ε_{A,X} ∶ LRA(X) → A(X)$ is an isomorphism
  for any non empty set $X$, for any $A$ in $[𝐒𝐞𝐭, 𝐒𝐞𝐭]_{ω⁺}$.
  Let us decompose the adjunction $L ⊣ R$ as follows.
  \[
  \begin{tikzpicture}
\draw[white,-,curve={ratio=-0.2}, line width=0.20833333333333334em] (22.265625em,-13.4375em) .. controls (24.609375em,-12.5em) and (26.953125em,-12.5em) .. (29.296875em,-13.4375em);
\draw[black,->, curve={ratio=-0.2}, ] (22.265625em,-13.4375em) .. controls (24.609375em,-12.5em) and (26.953125em,-12.5em) .. (29.296875em,-13.4375em);
\draw[white,-,curve={ratio=-0.2}, line width=0.20833333333333334em] (29.296875em,-15.208333333333334em) .. controls (26.953125em,-16.145833333333336em) and (24.609375em,-16.145833333333336em) .. (22.265625em,-15.208333333333334em);
\draw[black,->, curve={ratio=-0.2}, ] (29.296875em,-15.208333333333334em) .. controls (26.953125em,-16.145833333333336em) and (24.609375em,-16.145833333333336em) .. (22.265625em,-15.208333333333334em);
\draw[white,-,curve={ratio=-0.2}, line width=0.20833333333333334em] (33.723958333333336em,-13.4375em) .. controls (36.067708333333336em,-12.5em) and (38.411458333333336em,-12.5em) .. (40.755208333333336em,-13.4375em);
\draw[black,->, curve={ratio=-0.2}, ] (33.723958333333336em,-13.4375em) .. controls (36.067708333333336em,-12.5em) and (38.411458333333336em,-12.5em) .. (40.755208333333336em,-13.4375em);
\draw[white,-,curve={ratio=-0.2}, line width=0.20833333333333334em] (40.755208333333336em,-15.208333333333334em) .. controls (38.411458333333336em,-16.145833333333336em) and (36.067708333333336em,-16.145833333333336em) .. (33.723958333333336em,-15.208333333333334em);
\draw[black,->, curve={ratio=-0.2}, ] (40.755208333333336em,-15.208333333333334em) .. controls (38.411458333333336em,-16.145833333333336em) and (36.067708333333336em,-16.145833333333336em) .. (33.723958333333336em,-15.208333333333334em);
\draw[white,draw=none, line width=0.20833333333333334em] (25.78125em,-15.390625em) .. controls (25.78125em,-14.678819444444445em) and (25.78125em,-13.967013888888888em) .. (25.78125em,-13.255208333333334em);
\draw[black,draw=none, ] (25.78125em,-15.390625em) .. controls (25.78125em,-14.678819444444445em) and (25.78125em,-13.967013888888888em) .. (25.78125em,-13.255208333333334em);
\draw[white,draw=none, line width=0.20833333333333334em] (37.239583333333336em,-15.390625em) .. controls (37.239583333333336em,-14.678819444444445em) and (37.239583333333336em,-13.967013888888888em) .. (37.239583333333336em,-13.255208333333334em);
\draw[black,draw=none, ] (37.239583333333336em,-15.390625em) .. controls (37.239583333333336em,-14.678819444444445em) and (37.239583333333336em,-13.967013888888888em) .. (37.239583333333336em,-13.255208333333334em);
\node at (20.052083333333336em,-14.322916666666668em) {$[\BN, 𝐒𝐞𝐭]$} ;
\node at (31.510416666666668em,-14.322916666666668em) {$[𝐒𝐞𝐭^*, 𝐒𝐞𝐭]$} ;
\node at (42.96875em,-14.322916666666668em) {$[𝐒𝐞𝐭, 𝐒𝐞𝐭]$} ;
\node[scale=0.7] at (25.78125em,-12.208193883289955em) {$L_1$} ;
\node[scale=0.7] at (25.78125em,-16.42942074418985em) {$R_1$} ;
\node[scale=0.7] at (37.239583333333336em,-12.208193883289955em) {$L_2$} ;
\node[scale=0.7] at (37.239583333333336em,-16.42942074418985em) {$R_2 $} ;
\node[rotate=90,scale=0.7] at (25.78125em,-14.322916666666668em) {$\vdash$} ;
\node[rotate=90,scale=0.7] at (37.239583333333336em,-14.322916666666668em) {$\vdash$} ;
\end{tikzpicture}
  \]
  Therefore, the counit $LRA → A$ decomposes as
  \begin{equation}
    \label{eq:counit-decomposition}
    L_2 L_1 R_1 R_2 A \xrightarrow{L_2 \varepsilon_{1,A} R_2 A}
    L_2 R_2 A \xrightarrow{\varepsilon_{2,A}} A,
  \end{equation}
  and the desired property is that the image by $R_2$ of this morphism is an isomorphism.
  But the image of $R_2$ of the second morphism in \cref{eq:counit-decomposition}
  is an isomorphism since $L_2 ⊣ R_2$ is a \kl{coreflection}.
  Therefore, it is enough to show that
  $ L_1 R_1 R_2 A \xrightarrow{L_2 \varepsilon_{1,A} R_2 A} R_2 A $ is an isomorphism.

  We check that the essential image of the fully faithful functor
  $[𝐒𝐞𝐭^*, 𝐒𝐞𝐭]_{ω⁺} ↪ [𝐒𝐞𝐭^*, 𝐒𝐞𝐭] \xrightarrow{\Lan} [𝐒𝐞𝐭,𝐒𝐞𝐭]$ is $[𝐒𝐞𝐭, 𝐒𝐞𝐭]_{ω⁺, cc₀}$
  so that the following square is a weak pullback.
  \[
    \begin{tikzcd}
      {[𝐒𝐞𝐭^*, 𝐒𝐞𝐭]}_{ω⁺} \ar[r] \ar[d] &
      {[𝐒𝐞𝐭, 𝐒𝐞𝐭]_{ω⁺, cc₀}} \ar[d] \\
      {[{𝐒𝐞𝐭^*},𝐒𝐞𝐭]}  \ar[r, "\Lan", swap] & {[𝐒𝐞𝐭, 𝐒𝐞𝐭]}
    \end{tikzcd}
  \]
  Therefore, we get a functor $[\BN, 𝐒𝐞𝐭] →[𝐒𝐞𝐭^*, 𝐒𝐞𝐭]_{ω⁺}$, and it is actually
  an equivalence because the top functor above is an equivalence, and
  $[\BN, 𝐒𝐞𝐭] →[𝐒𝐞𝐭, 𝐒𝐞𝐭]_{ω⁺,cc₀}$ is an equivalence, which factors $L_1$,
  which means that for any $X$ in $[𝐒𝐞𝐭^*, 𝐒𝐞𝐭]_{ω⁺}$, the counit component
  $L_1 R_1 X → X$ is an isomorphism.
  Now, if $A$ is in $[𝐒𝐞𝐭, 𝐒𝐞𝐭]_{ω⁺}$, then $R_2 A$ is in $[𝐒𝐞𝐭^*, 𝐒𝐞𝐭]_{ω⁺}$,
  and therefore $L_1 R_1 R_2 A → R_2 A$ is an isomorphism as desired.
\end{proof}

\section{Application: A Generalized Recursion Principle for Simply-Typed Languages}
\label{sec:app-generalized-rec}

Initial semantics represent languages as initial models over a base monoidal
category $C$, which models contexts.
Consequently, the base category depends on the type system of the language considered.
For instance, for a simply-typed languages with type system $T$, one possible base
category is $[\Set^T,\Set^T]$ as in \cite[Chapter 6]{ZsidoPhd10}, see
\cite{FioreHur} or \cite[Section 7]{dblmcs} for alternative choices.
There is an adequate notion of \kl{simply-typed binding signature} and the initial
model can be understood as the syntax, in the spirit of initial semantics.
However, the recursion principle provided by initiality is limited to models
over the same base monoidal category, and in particular over the same type system.
This means we cannot directly translate between languages with different type systems by initiality.
In this section, we explain how to deal with this issue, exploiting the 1-functoriality of the categories of models.

\subsection{Models from Simply-Typed Signatures, 1-Functorially}
\label{subsec:extended-models-simply-typed}

A \kl{simply-typed binding signature} is a standard notion of signature for
simply-typed languages featuring variable binding, e.g., see \cite{FioreHur}.
Given a type system $T$, binding arities represent constructors by specifying the
output type $τ$ and the types of the arguments $t_i$ together with the types
$\vec{u_i}$ of the variables bound in each of them.
A language is then represented by a family of arities.

\begin{definition}
  A \AP\intro{simply-typed binding arity} over a set $T$ is an element of
  $((\vec{u_1}, t_1), …,(\vec{u_n}, t_n),τ) ∈ (T^* × T)^* × T$ which we write
  \[
  t_1^{(\vec{u_1})} × … × t_n^{(\vec{u_n})} → τ
  \]
\end{definition}

\begin{definition}
  A \AP\intro{simply-typed binding signature} over a set $T$ is a family
  $(αᵢ)_{i ∈ I}$ of of simply-typed binding arities $αᵢ$ over $T$.
\end{definition}

\begin{example}
  \label{ex:simply-typed-lambda-bindingsig}
  Let $T$ be a set with a binary operation $(⇒) ∶ T × T → T$ and an interpretation
  $g ∶ B → T$ of a set $B$ of \emph{base types}.
  The simply-typed lambda calculus on $T$ can be specified by the
  \kl{simply-typed binding signature} consisting of the following binding
  arities for each pair $(s,t) ∈ T^2$.
  \begin{align*}
    \app_{s,t} : (t ⇒ s) × t → s && \abs_{s,t} : s^{(t)} → (t ⇒ s)
  \end{align*}
\end{example}

This definition naturally assembles into a category of simply-typed signatures,
by requiring morphisms to translate the type system and labels while respecting arities.
An important operation on arities and signatures is \emph{retyping}:

\begin{definition}
  Given a \kl{simply-typed binding arity}
  $α = \left(t_1^{(\vec{u_1})} × … × t_n^{(\vec{u_n})} → τ \right)$ over $T$ and a function $g ∶T → T'$,
  the retyping $g^* α$ of $\alpha$ along $g$  is defined by applying $g$ to all the involved types:
  \[ g^* α := \left(g(t_1)^{(\vec{g(u_1)})} × … × g(t_n)^{(\vec{g(u_n)})} → g(τ) \right) \]
  The retyping of a simply-typed signature is then the retyping of each arity,
  i.e. $g^* (αᵢ)_{i ∈ I} = (g^* α_{i})_{i ∈ I}$.
\end{definition}

\begin{definition}
  We define the category \AP$\intro*\STSigCat$ of simply-typed binding signatures as follows.
  Objects are pairs of a set $T$ and a \kl{simply-typed binding signature} $S$ over $T$.
  A morphism between $(T,(αᵢ)_{i ∈ I})$ and $(T',(α'_{i'})_{i' ∈ I'} )$ consists of
  a pair of functions $(T \xrightarrow{g} T',  I \xrightarrow{h} I')$ translating the types
  and the labels in a way that is compatible with the arities, that is, such that
  $α'_{h(i)} = g^* α_i$ for all $i ∈ I$.
\end{definition}

The crucial results for our purpose are the following.

\begin{theorem}
  \label{thm:ModSig-STSig}
  There is a 1-functor $\STSigCat^{op} → \ModSig$, where the target 2-category is
  considered as a 1-category, mapping a \kl{simply-typed binding signature}
  $(αᵢ)_{i ∈ I}$ over $T$ to the following \kl{module signature} on $[𝐒𝐞𝐭ᵀ, 𝐒𝐞𝐭ᵀ]$.
  \[
    R ↦
    ∐_{{ αᵢ = \left(t_1^{(\vec{u_1})} × … × t_n^{(\vec{u_n})} → τ \right)_{i ∈ I} }}
    (R^{(\vec{u}₁)}_{t₁} × … × R^{(\vec{u_n})}_{t_{n}}) ⋅ 𝐲_{τ}
  \]
  where
  \begin{itemize}
    \item \AP $\intro*𝐲_{τ}∶T → 𝐒𝐞𝐭$ maps $t$ to $1$ if $t = τ$, and to $∅$ otherwise.
    \item $R^{(u_1, …, u_n)} ∶ 𝐒𝐞𝐭ᵀ → 𝐒𝐞𝐭ᵀ$ maps $X ∶ T → 𝐒𝐞𝐭$ to $R(X + 𝐲_{ u_1} + … + 𝐲_{u_n}) ∶ T → 𝐒𝐞𝐭$.
  \end{itemize}
\end{theorem}

\begin{theorem}
  \label{def:Models-STSig}
  We define the 1-functor \AP $\intro*\STSigModel$ as the composition
  \[ \STSigCat^{op} → \ModSig \xrightarrow{\Model} \Cat. \]
  Given a simply-typed signature, it computes its category of models,
  which always has an initial object;
  indeed, \citet[Chapter 6]{ZsidoPhd10} constructs an initial model for any simply-typed binding signature.
\end{theorem}

As a consequence, any morphism between \kl{simply-typed binding signatures}
$(g,h) ∶ (T,S) → (T',S')$ induces a functor
$\STSigModel(g,h)∶ \STSigModel(T',S') → \STSigModel(T,S)$
between their categories of models.
We get a unique morphism from the initial model of  $\STSigModel(T,S)$
to $\STSigModel(g,h)(M)$ for any model $M$ of $(T',S')$: this enables us to translate
a syntax into another one with a different type system.

\begin{proposition}
  \label{prop:extended-models-action}
  Given a morphism $(g,h) ∶ (T,S) → (T',S')$ between \kl{simply-typed binding
  signatures}, the functor $\STSigModel(g,h)$ maps a model $M$ of $(T',S')$ to
  the model $(g,h)^*M$ of $(T,S)$ whose underlying monad
  maps $X ∶ T → 𝐒𝐞𝐭$ to
  \[ τ ↦ M\left(∐_{t}X_t ⋅ 𝐲_{g(t)}\right)_{g(τ)} \]
\end{proposition}
\begin{proof}
  The functor $\STSigCat → \ModSig$ maps $(g,h)$ to
  a morphism of module signatures whose underlying
  lax monoidal functor is
  \[
  [𝐒𝐞𝐭ᵀ,𝐒𝐞𝐭ᵀ] \xrightarrow{[\Lan_g, 𝐒𝐞𝐭^g]} [𝐒𝐞𝐭^{T'},𝐒𝐞𝐭^{T'}].
  \]
  Unfolding $\Lan_g$ yields the claimed result.
\end{proof}
\begin{example}
  \label{ex:extended-initiality-weird}
  Consider the simply-typed binding signature $S$ of simply-typed lambda-calculus
  for the set $T$ generated by a set of base types $B$ and a binary operation $f ∶ T × T → T$.
  Now, consider $T'$ which is the disjoint union of  $T$ with a singleton set  $\{ τ \}$.
  The initial model of $(T',S)$ is like the simply-typed lambda calculus, but extended with a new type $τ$
  which cannot occur in an arrow type.
  We can obviously still embed the lambda calculus into this extended language: this is
  actually given by the initial morphism from the initial model of $(T,S)$
  to $(g,h)^*M'$, where $M'$ is the initial model of $(T',S)$,
  and $g ∶ T → T'$ is the inclusion and $h$ is the identity function.
  Note that in this particular case, $(g,h)^*M'$ is actually isomorphic to $M$.
\end{example}

\subsection{Generalized Recursion Principles}

\subsubsection{Extended Models of Simply-Typed Binding Signatures}

In the situation described above, we exploit a morphism $(g,h) ∶ (T,S) → (T', S')$
to turn a model $M$ of $(T',S')$ into a model $(g,h)^* M$ of $(T,S)$, and then get an initial
morphism from the initial model of $(T,S)$ to this model.
In short, $(g,h)$ allows us to see models of $(T',S')$ as models (in an extended sense) of $(T,S)$.
This suggests the following definition.

\begin{definition}
  We define the category \AP $\intro*\ExtModels{T,S}$ of \AP\intro{extended models} of $(T,S)$ as follows.
  An object $((T,S) \xrightarrow{g,h}(T',S'),M)$ is a model $M$ of a
  simply-typed binding signature $S'$ over a set $T'$ together with a morphism
  $(g,h)$ between $(T,S)$ and $(T',S')$.
  A morphism between extended models
  $((T,S) \xrightarrow{g_1,h_1}(T_1,S_1),M_1)$ and $((T,S) \xrightarrow{g_2,h_2}(T_2,S_2), M_2)$ consists of
  \begin{itemize}
    \item a morphism $(T_1,S_1)\xrightarrow{(g,h)} (T_2, S_2)$ such that
    $(g,h) ∘ (g_1, h_1) = (g_2, h_2)$;
    \item  a morphism
    $M_1 → (g,h)^*(M_2)$ in $\STSigModel(T_1,S_1)$.
  \end{itemize}
\end{definition}

\begin{remark}
  \label{rem:extmodels-grothendieck-construction}
  This is precisely
  the total category corresponding to the functor
  $((T,S)/\STSigCat)^{\op} → \STSigCat^{\op} → \Cat$ by
  the \emph{Grothendieck construction}~\cite[Definition 1.10.1]{Jacobs},
  where $(T,S)/\STSigCat$ denotes the coslice category under $(T,S)$:
  objects are pairs $(T',S')$ with a morphism $(g,h) ∶ (T,S) → (T',S')$,
  and morphisms are commutative triangles,
  and the projection functor $(T,S)/\STSigCat$ maps an object as above
  to $(T', S')$.

  Note that the Grothendieck construction of a functor $F ∶ C^{\op} →\Cat$
  yields a category whose objects are pairs of an object $c$ of $C$ and an object $x$ of $F(c)$;
   a morphism between $(c,x)$ and $(c',x')$ consists of
  a morphism $c \xrightarrow{f} c'$ and a morphism $x → F(f)(x')$.
\end{remark}

It is then possible to prove that $(T,S)$ is initial in this extended category of
models, in which translations between languages over different type systems is possible.
In summary, this provides simply-typed syntax with a generalized recursion principle, which
we illustrate in \cref{subsec:translating-pcf}.

\begin{theorem}
  \label{thm:initiality-extended}
  Let $M$ be the initial model $M$ of $(T,S)$.
  Then $((T,S) \xrightarrow{\id} (T,S), M)$ is initial in $\ExtModels{T,S}$.
\end{theorem}

This follows from the following general lemma, exploiting the fact that $\ExtModels{T,S}$
is a Grothendieck construction (\cref{rem:extmodels-grothendieck-construction}).

\begin{lemma}[{\cite[Lemma 6.11]{AHLM19}}]
  \label{lem:initiality-grothendieck}
  Let $F ∶ S^{op} → \Cat$ be a functor from a category with an initial object $0$, such that $F0$
  has an initial object $0'$. Then, the pair $(0,0')$ is initial
  in the total category of the fibration on $S$ generated by $F$ by the Grothendieck construction.
\end{lemma}

\subsubsection{Recovering Ahrens's Framework as a Grothendieck Construction}

We end this section by revisiting Ahrens' extended models with its initiality
property in light of \cref{thm:initiality-extended}.
Ahrens~\cite{ExtendedInitiality12} defines a notion of \emph{typed signature}
with an associated category of models~\cite[Definition 3.30]{ExtendedInitiality12}
that assembles monads on $𝐒𝐞𝐭ᵀ$ for different $T$.
A typed signature~\cite[Definition 3.25]{ExtendedInitiality12} consists of a
\emph{signature for types} $Σ$ and a \emph{term-signature}
$S$ over $Σ$ (see~\cite[3.1]{ExtendedInitiality12} and~\cite[Definition 3.24]{ExtendedInitiality12}).
First, any signature for types $Σ$ induces a category of \emph{algebras} $𝐒𝐞𝐭_{Σ}$: they are sets
equipped with an interpretation of the type system and one can show that there is an initial object
(it is the category of algebras of a finitary endofunctor on sets).
Then, a \emph{term-signature} $S$ over $Σ$
induces a functor $𝐒𝐞𝐭_{Σ} → \STSigCat$ mapping $T$ to a binding signature $S_T$ over $T$

\begin{example}
  Given a set of \emph{base types} $B$, the category of algebras of the type
  signature $Σ$ of the simply-typed lambda calculus are sets $T$ equipped with a
  binary operation $(⇒) ∶ T × T → T$ and a function $g ∶ B → T$.

  The term-signature for the simply-typed lambda calculus generates the functor
  $𝐒𝐞𝐭_{Σ} → \STSigCat$ that maps an algebra $(T,⇒,g)$ to the simply-typed
  binding signature given in \cref{ex:simply-typed-lambda-bindingsig}.
\end{example}

Ahrens' category of models of a typed signature is recovered by computing the
Grothendieck construction as for $\ExtModels{T,S}$: a model consists of a $Σ$-algebra
$T$ and a model $M$ of the binding signature $S_T$.
The existence of the initial model is again guaranteed by \cref{lem:initiality-grothendieck}.

\begin{proposition}
  \label{prop:revisiting-ahrens}
  Any model $(T, M)$ in the sense of Ahrens induces an \kl{extended model}
  $((T_0,S_{T_0}) \xrightarrow{i} (T,S), M)$, where $i$ is the image of the
  initial morphism $T_0 → T$ by the functor $𝐒𝐞𝐭_{Σ} → \STSigCat$.
  Moreover, this assignement is functorial.
\end{proposition}
\begin{proof}
  The functor $𝐒𝐞𝐭_{Σ} → \STSigCat$ factors as $𝐒𝐞𝐭_{Σ} → (T₀,S_{T_0})/\STSigCat → \STSigCat$,
  where $T_0$ is the initial object of $𝐒𝐞𝐭_{Σ}$, and the first functor maps
  $T$ to the image of the initial morphism $T_0 → T$.

  This induces a functor between the total categories built
  using the Grothendieck construction by \cite[Theorem 1.10.7]{Jacobs}.
\end{proof}

However, the converse does not hold: our category of \kl{extended models} is
\emph{larger} than Ahrens' category of models, as shown by the following
example.

\begin{example}
  \cref{ex:extended-initiality-weird} is an \kl{extended model} that is not a model
  of the simply-typed lambda calculus in the sense of Ahrens'.
  Indeed, $T'$ is not a model of $Σ$, because the arrow construction on types is
  partial: it is not defined when one of the arguments is $τ$. Even if we
  artificially extend the domain of the arrow construction on types, the initial model of
  $(T',S)$ lacks application and abstraction for $τ$.
\end{example}

\subsection{Application: Translating PCF to the Untyped Lambda Calculus}
\label{subsec:translating-pcf}
Ahrens illustrated his initiality result with the translations
from classical logic to intuitionistic logic~\cite[Section 4]{ExtendedInitiality12}
and from PCF to the untyped lambda calculus~\cite[Section 5]{ExtendedInitiality12}.
Since our category of \kl{extended models} is larger, these examples are still valid.
In this section, we detail the translation of PCF~\cite{plotkin1977lcf}, following
 Ahrens' presentation of the calculus.

\begin{definition}[Binding signature of PCF]
  The set $T_{PCF}$ of simple types is generated by two base types, $𝔹$ and $ℕ$,
  and a binary operation $-⇒- ∶ T_{PCF} × T_{PCF} → T_{PCF}$.

  The simply-typed binding signature $S_{PCF}$ of PCF consists of
  the family of binding arities for application and abstraction as in \cref{ex:simply-typed-lambda-bindingsig}, as well of the following binding arities (where we additionally label them for clarity).
  \begin{gather*}
    (\mathrm{true} ∶ () → 𝔹) \qquad
    (\mathrm{false} ∶ () → 𝔹)
    \\
    (\mathrm{0} ∶ () → ℕ) \qquad
    (\mathrm{succ} ∶ () → ℕ ⇒ ℕ)\qquad
    (\mathrm{pred} ∶ () → ℕ ⇒ ℕ)
    \\
    (\mathrm{Fix}_τ ∶ (τ ⇒ τ) → τ)_{τ ∈ T_{PCF}}
    \qquad
    (\mathrm{if}_{ι} ∶ () → 𝔹 ⇒ ι ⇒ ι ⇒ ι)_{ι ∈\{ 𝔹, ℕ \}}
  \end{gather*}
  Let $M_{PCF}$ denotes the initial model.
\end{definition}
\begin{definition}[Binding signature of untyped lambda calculus]
  The set $T_{ULC}$ consists of a single type $⋆$.
  The simply-typed binding signature $S_{ULC}$ consists of the following binding arities.
  \[
\abs ∶⋆^{(⋆)} → ⋆
  \qquad
  \app ∶ ⋆ × ⋆ → ⋆
  \]
  Let $M_{ULC}$ denotes the initial model.
  Exploiting the isomorphism $𝐒𝐞𝐭^{\{⋆\}} ≅ 𝐒𝐞𝐭$, we
  consider $M_{ULC}$ as a monad on $𝐒𝐞𝐭$.
\end{definition}
To translate PCF to the untyped lambda calculus, we need to give
$M_{ULC}$ a structure of \kl{extended model} for
$(T_{PCF}, S_{PCF})$.
That is, we need to find a simply-typed binding signature $S$ over a set $T$,
with a morphism $(g,h) ∶ (T_{PCF},S_{PCF}) → (S,T)$, and
a $(S,T)$-model structure on $M_{ULC}$.
An obvious choice is to take $T = T_{ULC}$ and $S = S_{ULC}$, since
$M_{ULC}$ is automatically a model of $(T,S)$.
We have indeed a function $g ∶ T_{PCF} → T$ mapping any type to $*$,
but unfortunately, there is no suitable function $h$, because for example
$S_{ULC}$ does not have any binding arity such as $\mathrm{0} ∶ () → g(ℕ)=⋆$.

Instead, we consider the binding signature $g^* S_{ULC}$ over $T_{PCF}$.
This means that we need to provide operations $1 → M_{ULC}$ for each constant
of $S_{ULC}$, a fix operation $M_{ULC} → M_{ULC}$.
Note that any closed term $t ∈ M_{ULC}(∅)$ induces an operation $1 → M_{ULC}$
whose $X$-component maps the unique element of $1$ to $M_{ULC}(i)(t)$, where $i ∶ ∅ → X$ is the initial function.
Exploiting the Church encoding of booleans, we pick
the closed term $λt.λf.t$ for $\mathrm{true}$, and $λt.λf.f$ for $\mathrm{false}$,
so that we can define the $\mathrm{if}$ constant as $λ b.λ t.λ f. b\ t\ f$.
Natural numbers can be similarly encoded. For the fixpoint operation,
we can define the operation $M_{ULC} → M_{ULC}$ as mapping $t$ to
$Y\ t$, where $Y$ is any fixpoint combinator, e.g.,
Curry's combinator $λf.(λx.f(x\ x))(λx.f(x\ x))$.

Let us make explicit what the initial morphism gives us.
We just showed that $M_{ULC}$ induces a model $M'_{ULC}$ of $g^* S_{ULC}$
with the same underlying monad. Initiality
provides us with a morphism $M_{PCF} → (g,h)^* M'_{ULC}$.
Unfolding the codomain using \cref{prop:extended-models-action},
we see that $(g,h)^* M'_{ULC}$ maps $Γ ∶ T_{PCF} → 𝐒𝐞𝐭$ and $τ ∈ T_{PCF}$ to
the set $M(∐_{t ∈ T_{ULC}}Γ_t)$. That is $(g,h)^*M'_{ULC}$ just forgets
the types of the variables in the input context $Γ$.
As a model morphism, the initial morphism translates all the PCF operations recursively
using the operations of $M'_{ULC}$ that we sketched above.

\section{Related Work}
\label{sec:related-work}

Different frameworks have been considered over time for initial semantics
\cite{DBLP:conf/lics/FiorePT99,DBLP:journals/iandc/HirschowitzM10,MATTHES2004155}.
\citet{lamiaux2025unifiedframeworkinitialsemantics} recently
presented a unified account of initial semantics, parametrized by a choice of a
base monoidal category on which we base this work.
We review briefly the relevant part of their work in \cref{sec:overview-initial-semantics}.
They have written an extensive overview of the different traditions
\cite[Section 6]{lamiaux2025unifiedframeworkinitialsemantics}, and we
refer to it for discussion on the links with others approaches.

Relating the framework of \citet{DBLP:conf/lics/FiorePT99} using $[\F,\Set]$
and that of \citet{hirscho:lam} using $[\Set,\Set]$ was first studied by
\citet{ZsidoPhd10} in her PhD dissertation.
In that work \cite[Chapter 4]{ZsidoPhd10}, \citeauthor{ZsidoPhd10} lifts the monoidal adjunction
between $[\F,\Set]$ and $[\Set,\Set]$ to the categories of monoids, then uses it
to directly build the initial model of $[\F,\Set]$ out of the one on $[\Set,\Set]$,
and vice versa.
We have generalized this result in \cref{sec:application-FSet}, by using
2-functoriality to lift the monoidal adjunction directly to models
(\cref{thm:lifting-models}), recovering Zsidó's result as a particular case
using that the adjunction is a coreflection (\cref{cor:zsido-preserves-initial}).
We have further refined it by factoring it through the equivalence $[\F,\Set] \simeq [\Set,\Set]_f$.

In her PhD, Zsidó also studied the links in the simply-typed case \cite[Chapter 7]{ZsidoPhd10}.
She lifted the adjunction of the respective monoidal categories by hand to
models --- which technically also gives her an adjunction in the untyped case,
though it is not mentioned --- but failed to conclude that initial objects were
preserved.\footnote{See the end of Section 1.2 of Zsidó's dissertation \cite{ZsidoPhd10}.}
We believe that this preservation could be proved following an argument
similar to that of the proof of \cref{cor:zsido-preserves-initial} -- that is,
given a coreflection, both adjoint functors preserve the initial object.
%

\citet{dblmcs} introduced
De Bruijn monads and modules over them and defined models of binding signatures.
We showed how they fit into our framework as monoids and modules in a \kl{binding-friendly monoidal} category.
Moreover, we deduce their equivalence between finitary De Bruijn models and
intersectional models in $[𝔽, 𝐒𝐞𝐭]$ from our 2-categorical machinery.

Generalizing the recursion principle of typed languages has only been studied for
simply-typed languages by \citet{ExtendedInitiality12}.
To do so, \citeauthor{ExtendedInitiality12} designed a specific notion of signatures and associated models to integrate the simply-typed
systems directly into it, fixing the shape of the base categories to be of the
form $[\Set^T,\Set^T]$ with $T$ an appropriate algebra for the type signature.
%
As discussed by \citet[Section 6.3.5]{lamiaux2025unifiedframeworkinitialsemantics},
the links with the usual monoidal setting was unclear.
In \cref{sec:app-generalized-rec}, we have used instead the 2-functoriality of models
to provide a generalized recursion principle based on the standard notion
of \kl{simply-typed binding signature}, without the need of introducing new signatures.
We showed that like ours, Ahrens' category of models can be recovered as
a total category by the Grothendieck construction; however
our recursion principle is more general.

Power and Tanaka~\cite{DBLP:journals/lisp/TanakaP06,10.1007/11417170_23} provide a general recipe to construct monoidal categories such as $[𝔽,𝐒𝐞𝐭]$ and some linear, simply-typed variants. Their recipe also yields a notion of (generalised) binding signature for each such monoidal category.


In the context of initial semantics, skew-monoidal categories have been used in particular by \citet{dblmcs}, \citet{FioreSzamozvancev}, and \citet{HoweTheorem}.
  As \cref{prop:debruijn-monads-monoids} suggests, \citet{dblmcs} could have focused on the monoidal category $[\BN, 𝐒𝐞𝐭]$ instead, at the cost of working with a more complicated monoidal product than theirs, which does not involve quotients. Similarly, \citet{FioreSzamozvancev} took advantage of their skew monoidal product to formalise their framework in a quotient-free type theory.
  Finally, the work by \citet{HoweTheorem} has since been superseded by that of \citet{CatFrameworkBisim} which works with truly monoidal categories.

\section{Conclusion}
\label{sec:conclusion}

In this work, we have shown that \kl{module signatures}, when parametrized by a
monoidal category, have a 2-categorical structure, and that in this case
the associated category of models can be computed by a 2-functor.
%
We have then leveraged this 2-functoriality to propagate adjunctions and equivalences
on the base monoidal categories, corresponding to the different implementations of
abstract syntax, to the associated categories of models.
This enabled us to recover the results presented by \citet{ZsidoPhd10} and
\citet{HHLM} using a generic proof method, whereas previously, these results only
had instance-specific proofs.
%
Using 1-functoriality of the 2-functoriality of models, we have also designed a
generalized recursion principle for simply-typed languages with variables binding.
This enables us to recover and understand the framework developed by
\citet{ExtendedInitiality12} as a total category, and to generalize it.
In both cases, we hope that the solutions we have designed using 2-functoriality
will scale to more complex languages with variable binding, such as polymorphic
languages like System F.

\subsection{Open Problems}

With our work, we open doors to comparing different implementations and
proving more general recursion principles.
While we have addressed some instances, numerous approaches remain to be
compared or generalized.

\begin{enumerate}
  \item In the untyped case, we have compared unscoped syntax with well-scoped syntax.
        It remains to understand how the nominal approach~\cite{PittsAM:newaas} fits into the picture,
        starting from Power's monoidal structure on the category of nominal sets~\cite[Theorem 4.6]{Power}.

  \item The 2-functor of \cref{thm:2-functoriality-bindingfriendly} only deals  with untyped syntax.
        It would be interesting to generalize $\BindMonCat$ to simply-typed languages
        so as to generalize our category of \kl{extended models} to other categories than $[𝐒𝐞𝐭^T, 𝐒𝐞𝐭^T]$
        which were considered by \citet{FioreHur} and \citet{dblmcs}.
  \item Several accounts of initial semantics for equations or reductions between
        terms have been given, using different notions of context
        \cite{FioreHurEquational,DBLP:journals/corr/abs-1107-5252,AHLM19}.
        It would be interesting to extend our 2-functor to incorporate
        equations or reductions, in order to compare these approaches.
\item Initial semantics results for polymorphic type systems like system F or F$_ω$ exist
      \cite{HamanaPoly,DBLP:conf/lics/FioreH13}, but are currently underdeveloped.
      One of the challenges is that the binding of type variables makes it
      necessary, formally, to consider changing the base category modelling contexts.
      We hope that our techniques will make it easier to construct useful
      initial semantics for polymorphic type systems, and scale to it.
\item Some recent accounts of initial semantics have been working on skew monoidal categories~\cite{dblmcs,HoweTheorem,FioreSzamozvancev}.
      As discussed in \cref{sec:related-work}, skew-monoidal categories
      can sometimes be avoided, and our framework directly applied.
      Nonetheless, we are not sure to which extent our constructions readily extend
      to this more general setting.
\end{enumerate}

\begin{acks}
 We thank Tom Hirschowitz and Thea Li for helpful discussions, and the anonymous referees for their detailed feedback.
This work received government funding managed by the French National Research Agency under the France 2030 program, reference “ANR-22-EXES-0013”.
\end{acks}

\appendix

\section{Proof of \texorpdfstring{\cref{thm:exponential-parammodule}}{Theorem~\ref{thm:exponential-parammodule}}}
\label{app:exponential-parammodule}

\begin{lemma}
  \label{lem:exponential-freealg}
  Let $C$ be a category with finite products and
  $T$ be a monad on $C$ preserving those products.
  Let $c$ be an exponentiable object of $C$, that is, such that
  given an algebra $TX → X$, the algebra structure on $X^c$
  is given by the transpose of the following morphism makes it
  the exponential $X^{Tc}$ in the category of $T$-algebras.
  \[ T(Xᶜ) × c → T(Xᶜ) × Tc
    ≅ T(Xᶜ × c) → T(X) → X
  \]
\end{lemma}
\begin{proof}
  \renewcommand{\yoneda}{\mathbf{y}}

  We exploit the following characterisation of the category of $T$-algebras, as the following pullback~\cite[Theorem 14]{Street72}.
  \[
    \begin{tikzcd}
       Alg(T) \ar[r] \ar[d] \ar[dr, phantom, very near start, "\lrcorner"] & {[\Kl(T)^{\op}, 𝐒𝐞𝐭]} \ar[d, "{[L^{\op}, 𝐒𝐞𝐭]}"] \\
        C \ar[r, "𝐲"'] & {[C^{\op}, 𝐒𝐞𝐭]}
    \end{tikzcd}
  \]
  where $𝐲$ denotes the Yoneda embedding, and $L ∶ C → \Kl(T)$ denotes the canonical left adjoint
  into the Kleisli category of $T$.

   Let us show that $ 𝐲_{Lc}$ is exponentiable in $[\Kl(T)^{\op}, 𝐒𝐞𝐭]$.
  Note that since $\Kl{T}$ is not small in general,
   $[\Kl{T}^{\op}, 𝐒𝐞𝐭]$ is not necessarily cartesian closed.
   However, it fully faithfully embeds into the category $[\Kl(T)^{\op}, \mathbf{SET}]$ where
   $\mathbf{SET}$ is the category of large sets\footnote{Assuming a set universe is not
   strictly necessary but makes the proof simpler.}, which is cartesian closed by replaying
   the proof of \cite[Proposition A.1.5.5]{Johnstone1} for the larger universe $\mathbf{SET}$.

   Let $A$ be a functor from $\Kl(T)^{\op}$ to $𝐒𝐞𝐭$.
   Let us check that $A^{𝐲_{Lc}}(X)$ is small, so that the exponential
   $A^{𝐲_{Lc}}$ in $[\Kl{T}^{\op}, \mathbf{SET}]$
    actually lives in $[\Kl{T}^{\op}, 𝐒𝐞𝐭]$.
   By the Yoneda Lemma, $A^{𝐲_{Lc}}$ maps $X$ to
   \begin{align}
     \hom_{[\Kl{T}^{\op}, \mathbf{SET}]}( 𝐲_X , A^{𝐲_{Lc}})
   & ≅ \hom_{[\Kl{T}^{\op}, \mathbf{SET}]}(𝐲_X × 𝐲_{Lc}, A) \notag
   \\
   & ≅ \hom_{[\Kl{T}^{\op}, \mathbf{SET}]}(𝐲_{X × Lc}, A) & \text{ By continuity of the Yoneda embedding}
   \notag
   \\ & ≅ A(X × Lc) & \text{By the Yoneda Lemma}
   \label{eq:small-expo}
  \end{align}
  which is indeed small.

  Now, let us assume that $A ∘ L^{\op} ≅ 𝐲_{d}$ for some object $d$ of $C$. Let us show
  that $A^{𝐲_{Lc}} ∘ L^{\op}$ is representable.
  \begin{align*}
    A^{𝐲_{Lc}} L X & ≅ A(L X × Lc) & \text{By \eqref{eq:small-expo}}
    \\ & ≅ A(L(X × c)) & \text{By preservation of products}
    \\ & ≅ \hom(X × c, d) &\text{By representability of $A$}
    \\ & ≅ \hom(X, d ^c)& \text{By exponentiability of $c$}
    \\ & = 𝐲_{d^c}(X)
  \end{align*}
  Let us explain why $L$ preserves products,
   Kleisli morphisms $A → T(X × Y)$ are in bijection with morphisms $A → T X × T Y$,
    and the latter are in bijection with pairs of Kleisli morphisms $A → T X$ and $A → T Y$.

    What we have shown is that $A^{𝐲_{Lc}}$ is in the pullback
    characterising the category of $T$-algebras, the underyling object
    of $C$ being $d^c$, if $d$ is the underlying object of $A$.
    Moreover, it is the really the exponential of $A$ by $Lc$ in
    $[\Kl(T)^{\op},\mathbf{SET}]$ in which $T$-algebras fully faithfully embeds.
    It follows that it is also the exponential in $T$-algebras.
\end{proof}

\begin{lemma}
  \label{lem:expo-cst-functors}
  Let $D$ be a category with finite products, $d$ an exponentiable object of $D$.
  Then given any category $C$, the constant functor mapping any object of $c$ to
  $d$ is also exponentiable, with $Fᵈ$ defined as $F(-)ᵈ$.
\end{lemma}
We are now ready to prove \cref{thm:exponential-parammodule}.

\begin{proof}[Proof of \cref{thm:exponential-parammodule}]

  By \cref{thm:monadic-param-modules}, the category of \kl{module signatures}
  is monadic over $[\Mon(C),C]$, and the monad $T$ maps $Σ$ to $T(Σ) := Σ(-)⊗-$.
  Because $-⊗-$ preserves binary products on the left, this monad preserves
  binary products.

  By Lemma~\ref{lem:expo-cst-functors},
  the endofunctor $- × I$ on $\Mon(C) → C$ has a right adjoint $R$ mapping $Σ$ to $Σ(-)ᴵ$.

  By Lemma~\ref{lem:exponential-freealg}, the exponential functor $R$
  lifts to the category of \kl{module signatures} as claimed.

  It is the exponential $Σ^Θ$ in the category of module signatures
   since the monad of \cref{thm:monadic-param-modules} maps $I$ to
   $Θ$ (up to isomorphism).
\end{proof}

\section{Proof of \texorpdfstring{\cref{prop:param-module-funct}}{Proposition~\ref{prop:param-module-funct}}}
\label{app:bindingfriendly-sig}

\begin{definition}
    Let \AP$\intro*\ModSigps$ be the sub-2-category of $\ModSig$ with 1-cells
    $(C,Σ) \xrightarrow{(F,α)} (D,Σ')$ such that $α$ is an isomorphism.
  \end{definition}
  \begin{definition}
    A \AP\intro{binding-friendly signature} is a 2-functor from $\BindMonCat$ to $\ModSigps$ commuting with
    the projection to the 2-category of monoidal categories,
    lax monoidal functors and monoidal transformations.
  \end{definition}

  \begin{example}
    The tautological binding-friendly signature maps a \kl{binding-friendly monoidal} category $C$
    to the module signature $Θ∶\Mon(C) → \Mod(C)$.
  \end{example}

  \Cref{prop:param-module-funct} follows from the fact that binding-friendly signatures are stable under finite products, non-empty coproducts, and exponential by the unit, as stated
  in the following lemma. It is crucial there that we restrict
  the definition of \kl{binding-friendly signature} to
  the sub-2-category $\ModSigps$ of $\ModSig$, otherwise those stability properties do not hold.
  \begin{lemma}
    \label{lem:bindfriendly-stable}
    \begin{enumerate}
      \item Let $(Σ_i)_{i ∈ I}$ be a finite family of \kl{binding-friendly signatures}.
      Then their pointwise product $ΠᵢΣᵢ$ defined as $C ↦∏ᵢΣᵢC$ is a binding-friendly signature.
      \item Let $(Σ_i)_{i ∈ I}$ be a non-empty family of \kl{binding-friendly signatures}.
    Then their pointwise coproduct $∐ᵢΣᵢ$ defined as $C ↦∐ᵢΣᵢC$ is a binding-friendly signature.
    \item Let $Σ$ be a \kl{binding-friendly signature}.
    Then the pointwise exponential by unit $Σᴵ$ defined as $C ↦(ΣC)ᴵ$ is a binding-friendly signature.
    \end{enumerate}
  \end{lemma}

\section{Correspondence Between De Bruijn Modules and Modules over Monoids}

\label{app:debruijn-modules}
\cref{prop:debruijn-monads-monoids} states that the category of De Bruijn monads
is equivalent to the category of monoids in the monoidal category $[\BN, 𝐒𝐞𝐭]$.
In this section, we show that the category of De Bruijn modules~\cite[Definition 6.8]{dblmcs} is equivalent
to the category of modules over monoids in $[\BN, 𝐒𝐞𝐭]$.

\begin{definition}
    The category of De Bruijn modules over a De Bruijn monad $X$ is the category
    of modules relative to $X$ (see~\cite[Definition 14]{DBLP:journals/corr/abs-1107-5252}).
\end{definition}
\begin{proposition}
    The category of modules relative to a monad $X$  relative to some $J ∶ ℂ → 𝔻$ is
    isomorphic to the category from the
    Kleisli category~\cite[2.3]{DBLP:journals/corr/AltenkirchCU14} of $R$ to $𝔻$.
\end{proposition}
\begin{proof}
    Immediate, by unfolding the definitions.
\end{proof}
The main theorem is the following.
\begin{theorem}
    \label{thm:relmod-module}
      Let $X$ be a relative monad along some functor $J ∶ ℂ → 𝔻$
      such that the pointwise left Kan extension $\Lan_J ∶ [ℂ, 𝔻] → [𝔻, 𝔻]$ exists.
       Then, the category of relative modules over $X$,
       is isomorphic to the category
      of modules over $X$, i.e., algebras for the monad
      $-⊗ X$, for the skew monoidal structure on $[ℂ, 𝔻]$ induced by
      \cite[Theorem 3.1]{DBLP:journals/corr/AltenkirchCU14}.
    \end{theorem}
    \begin{theorem}
        The previous isomorphisms of categories for each relative monad $X$ gather into
        an isomorphism of categories between
        the total category of relative modules
        and the total category of modules over monoids in $[ℂ, 𝔻]$.
    \end{theorem}
    \begin{corollary}
        The category of De Bruijn modules is isomorphic to the category of modules over monoids in $[\BN, 𝐒𝐞𝐭]$.
    \end{corollary}
    In the rest of this section we focus on the proof of \cref{thm:relmod-module}.
    This follows from the following general lemmas.
    \begin{lemma}
    Let $L∶ 𝒜 → ℬ$ be a bijective-on-objects functor and $𝒞$ be category such that the left Kan
    extension $\Lan_L ∶ [𝒜,𝒞] → [ℬ,𝒞]$ exists. Then, the precomposition functor $-∘L
    ∶  [ℬ,𝒞] → [𝒜,𝒞]$ is strictly monadic.
    \end{lemma}
    \begin{proof}
      We apply Beck monadicity theorem~\cite[Theorem VI.7.1]{MacLane:cwm}.
      It is enough to show that the precomposition functor $L^* = - ∘ L$ creates
      pointwise colimits, since absolute coequalisers are pointwise (they are
      preserved by evaluation functors).

      Note that given a functor $F∶𝒜→ℬ$ and a diagram $d∶ J → 𝒜$, if there exists
      an isomorphism $ ColimCocones(d) ≅ ColimCocones(F∘ d)$
       such that the following diagram commutes and is a pullback, where $Cocones(G)$
       is the category of cocones over $G$ and $ColimCocones(G)$ is the full
       subcategory of initial objects therein, then the functor $F$ creates the
       $d$-colimits, and this is a necessary condition in case the colimit $F∘d$ exists.
    \[\begin{tikzcd}
     {ColimCocones(d)} & {ColimCocones(F∘d)} \\
     {Cocones(d)} & {Cocones(F∘d)}
     \arrow[from=1-1, to=2-1]
     \arrow[from=2-1, to=2-2]
     \arrow[from=1-2, to=2-2]
     \arrow["\cong", dashed, from=1-1, to=1-2]
    \end{tikzcd}\]

       Assume a diagram $d ∶ 𝒟 → [ℬ, 𝒞]$ such that the colimit of $L^*∘d$ is
       computed pointwise. Denoting $K ∶ |𝒜| → 𝒜$ the discrete category inclusion,
       this means that the colimit $K^* ∘ L^* ∘ d$ exist, and thus,
       by~\cite[Theorem V.3.2]{MacLane:cwm}, the plain squares in the below diagram are pullbacks
       with noted isomorphisms
    \[\begin{tikzcd}
     & {ColimCocones(K^*∘L^*∘d)} \\
     {ColimCocones(d)} && {ColimCocones(L^*∘d)} \\
     & {Cocones(K^*∘L^*∘d)} \\
     {Cocones(d)} && {Cocones(L^*∘d)}
     \arrow[from=2-1, to=4-1]
     \arrow[from=1-2, to=3-2]
     \arrow["\cong", from=2-1, to=1-2]
     \arrow[from=2-3, to=4-3]
     \arrow["{≅}"', from=2-3, to=1-2]
     \arrow[from=4-3, to=3-2]
     \arrow[from=4-1, to=4-3]
     \arrow[from=4-1, to=3-2]
     \arrow["{≅}"{pos=0.2}, dashed, from=2-1, to=2-3]
    \end{tikzcd}\]

    Therefore, we get the dashed arrow, which is an isomorphism by the cancellation
    property of isomorphisms. Moreover, by the pasting law for pullbacks~\cite[Exercise
    III.4.8]{MacLane:cwm}, the front square is a pullback, as desired.
    \end{proof}
    \begin{lemma}
      \label{lem:rel-adj-lan}
      Let $ ℂ(LA,B)≅𝔻(JA,UB)$ be a relative adjunction.
      If the pointwise left Kan extension
      $\Lan_J$ exists, then the (pointwise) left Kan extension
      $\Lan_L(-)$ can be computed as  $\Lan_J(-) ∘ U$.
    \end{lemma}
    \begin{proof}
      \begin{align*}
        \Lan_J F ∘ U (x)
        \\ & ≅ ∫^{c∈ℂ}𝔻(Jc, Ux)×Fc \tag{By definition of pointwise left Kan extension~\cite[§3.1]{DBLP:journals/corr/AltenkirchCU14}}
        \\ & ≅ ∫^{c∈ℂ}ℂ(Lc, x)×Fc \tag{By the relative adjunction}
        \\ & ≅ \Lan_L(F)(x) \tag{By definition of pointwise left Kan extension}
        \end{align*}
    \end{proof}
    \begin{proof}[Proof of \cref{thm:relmod-module}]
      Let $L$ and $U$ denote the free and forgetful functors $ℂ \xrightarrow{L}
      \Kl_X \xrightarrow{U} 𝔻$ inducing a relative adjunction
      \begin{align}
        \label{eq:relative-adj-kl}
        \Kl_X(LA,B)≅𝔻(JA,UB).
      \end{align}
      We can define back and forth functors between $[\Kl_X,
      𝔻]$ and $[ℂ,𝔻]$ defined as follows:
    \[\begin{tikzcd}
     {[\Kl_X,𝔻]} && {[ℂ,𝔻]} \\
     & {[𝔻,𝔻]}
     \arrow["{[L,𝔻]}", from=1-1, to=1-3]
     \arrow["{\Lan_J}", from=1-3, to=2-2]
     \arrow["{[U,𝔻]}", from=2-2, to=1-1]
    \end{tikzcd}\]
    By Lemma~\ref{lem:rel-adj-lan}, these functors are adjoint.
      It is straightforward to check that the induced monad is $-⊗ X$.

    \end{proof}

\bibliographystyle{ACM-Reference-Format}
\bibliography{bib}

\end{document}